\documentclass[lettersize,journal]{IEEEtran}
\usepackage{amsmath,amsfonts}
\usepackage{algorithmic}
\usepackage{algorithm}
\usepackage{array}
\usepackage[caption=false,font=normalsize,labelfont=rm,textfont=rm]{subfig}
\usepackage{textcomp}
\usepackage{stfloats}
\usepackage{url}
\usepackage{verbatim}
\usepackage{graphicx}
\usepackage{cite}
\usepackage{balance}
\usepackage{subcaption}
\usepackage{bm}
\usepackage{orcidlink}
\newtheorem{definition}{Definition}
\newtheorem{proposition}{Proposition}

\begin{document}

\title{Multi-hop Differential Topology based Algorithms \\ for Resilient Network of UAV Swarm}

\author{Huan Lin~\orcidlink{0009-0002-7217-6139}, Lianghui Ding~\orcidlink{0000-0002-3231-3613 },
\thanks{Huan Lin and Lianghui Ding are with the Institute of Image Communication and Network Engineering, Department of Electronic Engineering, School of Electronic Information and Electrical Engineering, Shanghai Jiao Tong University, Shanghai 200240, China. Email: lhzt715@sjtu.edu.cn; lhding@sjtu.edu.cn.}

}



\maketitle

\begin{abstract}
Unmanned aerial vehicle (UAV) swarm networks face severe challenges of communication network split (CNS) issues caused by massive damage in hostile environments. In this paper, we propose a new paradigm to restore network connectivity by repositioning remaining UAVs based on damage information within local topologies. Particularly, the locations of destroyed UAVs distributed in gaps between disconnected sub-nets are considered for recovery trajectory planning. Specifically, we construct the multi-hop differential sub-graph (MDSG) to represent local damage-varying topologies. Based on this, we develop two distinct algorithms to address CNS issues. The first approach leverages an artificial potential field algorithm to calculate the recovery velocities via MDSG, enabling simple deployment on low-intelligence UAVs. In the second approach, we design an MDSG-based graph convolution framework to find the recovery topology for high-intelligence swarms. As per the unique topology of MDSG, we propose a novel bipartite graph convolution operation, enhanced with a batch-processing mechanism to improve graph convolution efficiency.
Simulation results show that the proposed algorithms expedite the recovery with significant margin while improving the spatial coverage and topology degree uniformity after recovery.
\end{abstract}


\section{Introduction}
\IEEEPARstart{T}{echnology} of unmanned aerial vehicle (UAV) has developed rapidly in recent years and has been widely applied with the advantages of low cost and high flexibility. Meanwhile, to overcome the limited capability of single UAVs, the UAV swarm composed of hundreds of UAVs has been developed to adapt to dangerous environments and conduct various complex tasks, such as rescue \cite{rescue}, reconnaissance \cite{reconnaissance}, target attack \cite{attack}, etc. 
The UAV swarm usually forms an ad-hoc network for collaboration, where task and control information is transmitted between UAVs through wireless communication links. However, the UAV swarm network (USNET) is susceptible as the UAVs risk hostile strikes and external destructions. The damage of massive UAVs can result in substantial communication performance degradation and even divide the USNET into isolated sub-nets, causing the \textit{communication network split} (CNS) issue. Several pro-active resilient mechanisms were proposed to reduce the probability of CNS issues \cite{active1, active2, active3}, however, these strategies can not maintain the connectivity under massive damage scenarios. Therefore, the USNET requires reactive resilient algorithms to reconstitute the connectivity after the CNS issue.

The studies on resilient algorithms originated in wireless sensor networks (WSNs) under static scenarios \cite{wsn1, wsn2, wsn3, wsn4}. Subsequently, the mobility-based resilient algorithms were developed for mobile WSNs \cite{mv_cut1, mv_cut2, mv_cut3, mv_ms1, mv_wsn1, mv_wsn2, mv_wsn3, mv_opt1, mv_opt2, mv_opt3}, and were later extended for USNETs \cite{hero, sidr, mgc, demd}. The mobility-based algorithms can be divided into three categories. The first category of algorithms recovered the connectivity by repositioning the remaining nodes at the locations of destructed critical nodes \cite{mv_cut1, mv_cut2, mv_cut3, mv_ms1}, so-called the \textit{critical-node-based} algorithms. The \textit{meeting-point-based} algorithms as the second category aggregated the remaining toward a pre-defined point \cite{mv_wsn1, mv_wsn2, mv_wsn3, hero, sidr}. The third category, the \textit{optimization-based} algorithms, formulated the CNS recovery as an optimization problem and generated the solutions of recovery trajectories or locations for the remaining nodes \cite{mv_opt1, mv_opt2, mv_opt3, mgc, demd}. 

However, there are still several challenges in solving the CNS issues with mobility-based resilient algorithms. 
Firstly, many existing algorithms could not restore the connectivity under massive damage scenarios. For example, algorithms in \cite{mv_cut1, mv_cut2, mv_cut3, mv_ms1, mv_wsn1, mv_wsn2} were designed to handle CNS issues caused by single or multiple UAV destructions. Many heuristic algorithms \cite{mv_wsn3, mv_opt1, mv_opt2, mv_opt3, demd} were divergent and thus could not ensure the connectivity, especially more than 50\% of UAVs were damaged. 
The second challenge lies in the insufficient use of topology information. For example, the \textit{meeting-point-based} algorithms \cite{mv_wsn1, mv_wsn2, mv_wsn3, hero} disregarded the network topology completely, and the algorithms in \cite{mv_opt3, sidr, mgc} considered only one-hop neighbors that resulted in a limited receptive field. 
Thirdly, many existing algorithms \cite{mv_opt1, mv_opt2, mv_opt3, sidr, mgc} only focused on the remaining nodes but neglected the information of the damage itself. It is worth noting that the network is connected before the damage, hence the locations of destructed nodes distributed in the gaps between the divided sub-nets can guide the recovery trajectories. Other algorithms \cite{mv_cut1, mv_cut2, mv_cut3, mv_ms1} considered destructed nodes but required prolonged execution time to determine the critical nodes.


%
In this paper, we propose to restore the connectivity of the swarm networks by repositioning the remaining UAVs based on the locations of destroyed UAVs within local topologies. Specifically, we introduce the multi-hop differential sub-graph (MDSG) to represent the local damage-varying topologies, where each node's receptive field is extended into its multi-hop neighbors to capture additional features from local topology. Based on this, we develop two approaches to generate solutions for CNS issues utilizing MDSG. The first approach employs an artificial potential field algorithm to calculate the recovery velocities through MDSG, enabling simple deployment on low-intelligence UAVs. For the second approach, we design an MDSG-based graph convolution framework to determine recovery topology for high-intelligence swarms. As per the unique topology of MDSG, we propose a novel bipartite graph convolution operation, combined with a batch-processing mechanism to enhance the graph convolution performance. 
Simulation results indicate that the proposed algorithms expedite the connectivity recovery process. We also introduce the spatial coverage and the degree distribution to evaluate network topology, and our approaches showed significant improvements in recovered topology enhancement.
Our contributions in this paper are summarized as follows.
\begin{enumerate}
    \item We propose a new paradigm to reconstitute the connectivity of the swarm networks through damage information within local topologies. Specifically, we construct the multi-hop differential sub-graph (MDSG) to represent local damage-varying topologies.
    \item Based on MDSG, we develop an artificial potential field algorithm for simple deployment on low-intelligence UAVs. We also design an MDSG-based graph convolution framework for high-intelligence swarms, enhanced by a novel bipartite graph convolution operation and a batch-processing mechanism.
    \item We theoretically prove the convergence of the proposed algorithms, and simulation results show that the proposed algorithms can restore the connectivity more quickly while improving the spatial coverage and topology degree uniformity after recovery.
\end{enumerate}

The remainder of this paper is organized as follows. Section \uppercase\expandafter{\romannumeral2} analyzes the related work. Section \uppercase\expandafter{\romannumeral3} forms the system model of the USNET graph and CNS issue. Section \uppercase\expandafter{\romannumeral4} and section \uppercase\expandafter{\romannumeral5} introduce the proposed methods with the traditional algorithm and deep-learning algorithm, respectively, and experiment results are given in Section \uppercase\expandafter{\romannumeral6}. Finally, Section \uppercase\expandafter{\romannumeral7} presents the conclusion.

\begin{figure*}[!t]
\vspace{-1em}
\centerline{
\subfloat[CNS issue in USNET]{\includegraphics[width=.33\linewidth]{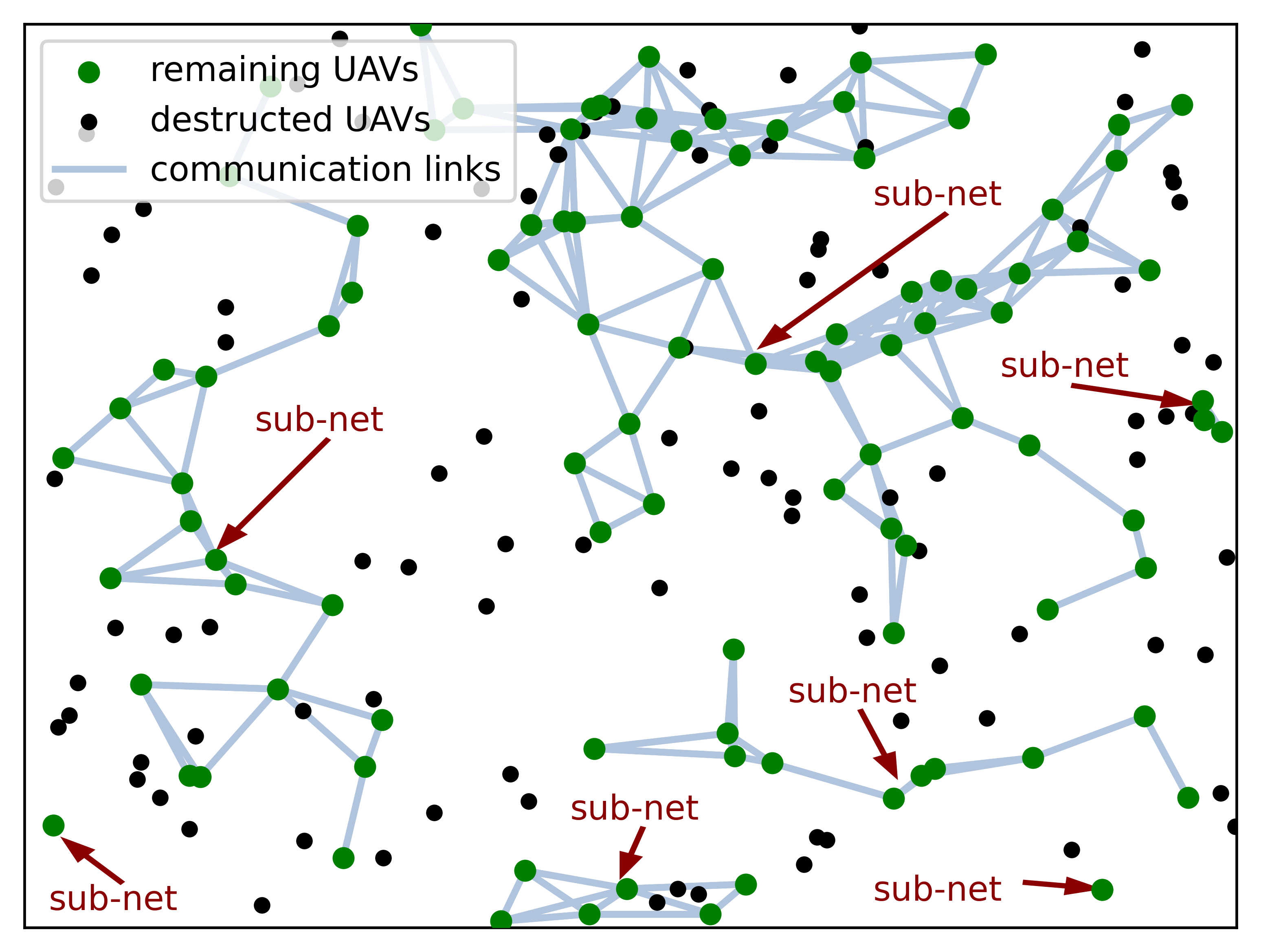}}
\subfloat[a meeting-point-based solution]{\includegraphics[width=.33\linewidth]{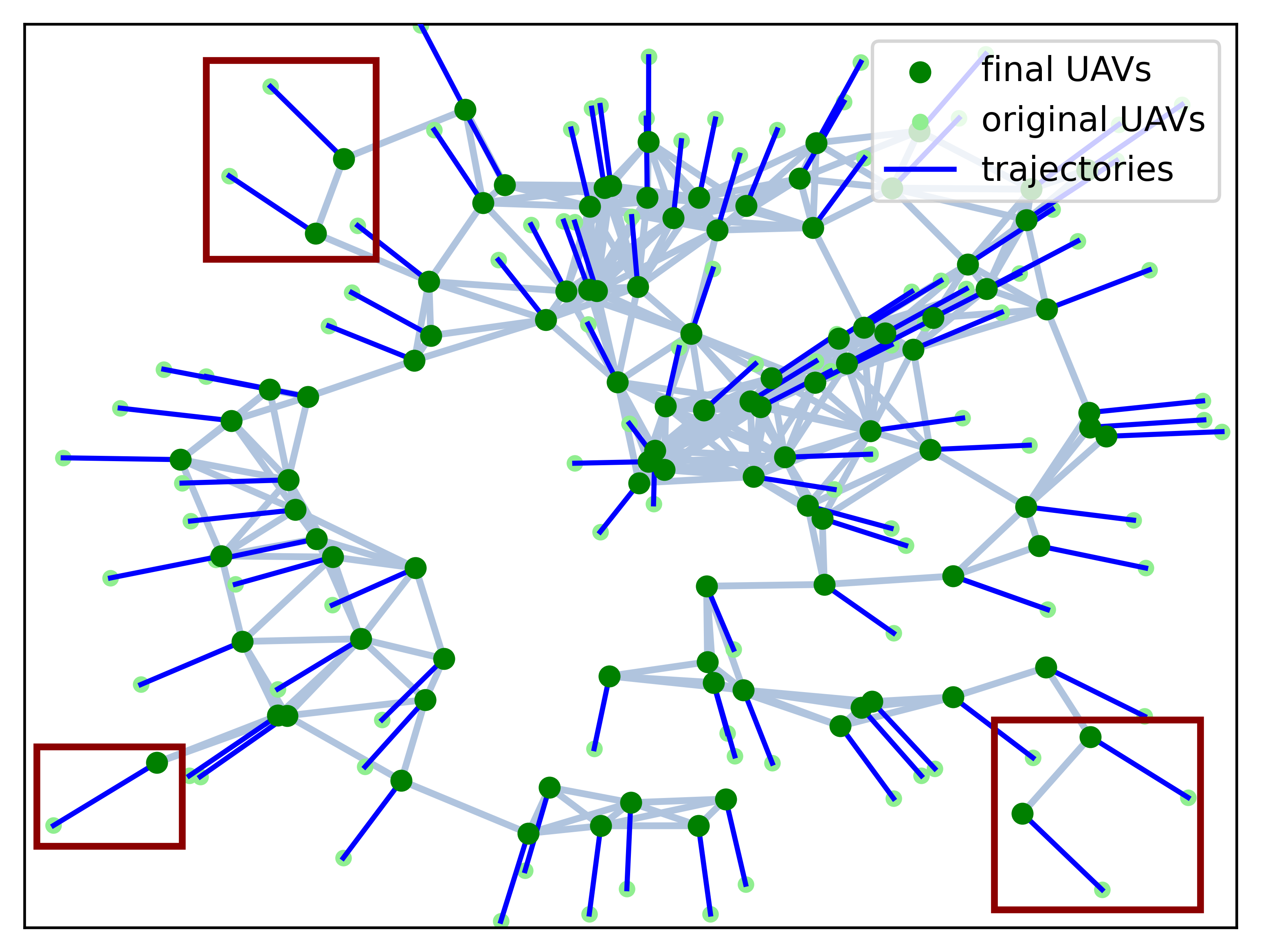}}
\subfloat[an optimization-based solution]{\includegraphics[width=.33\linewidth]{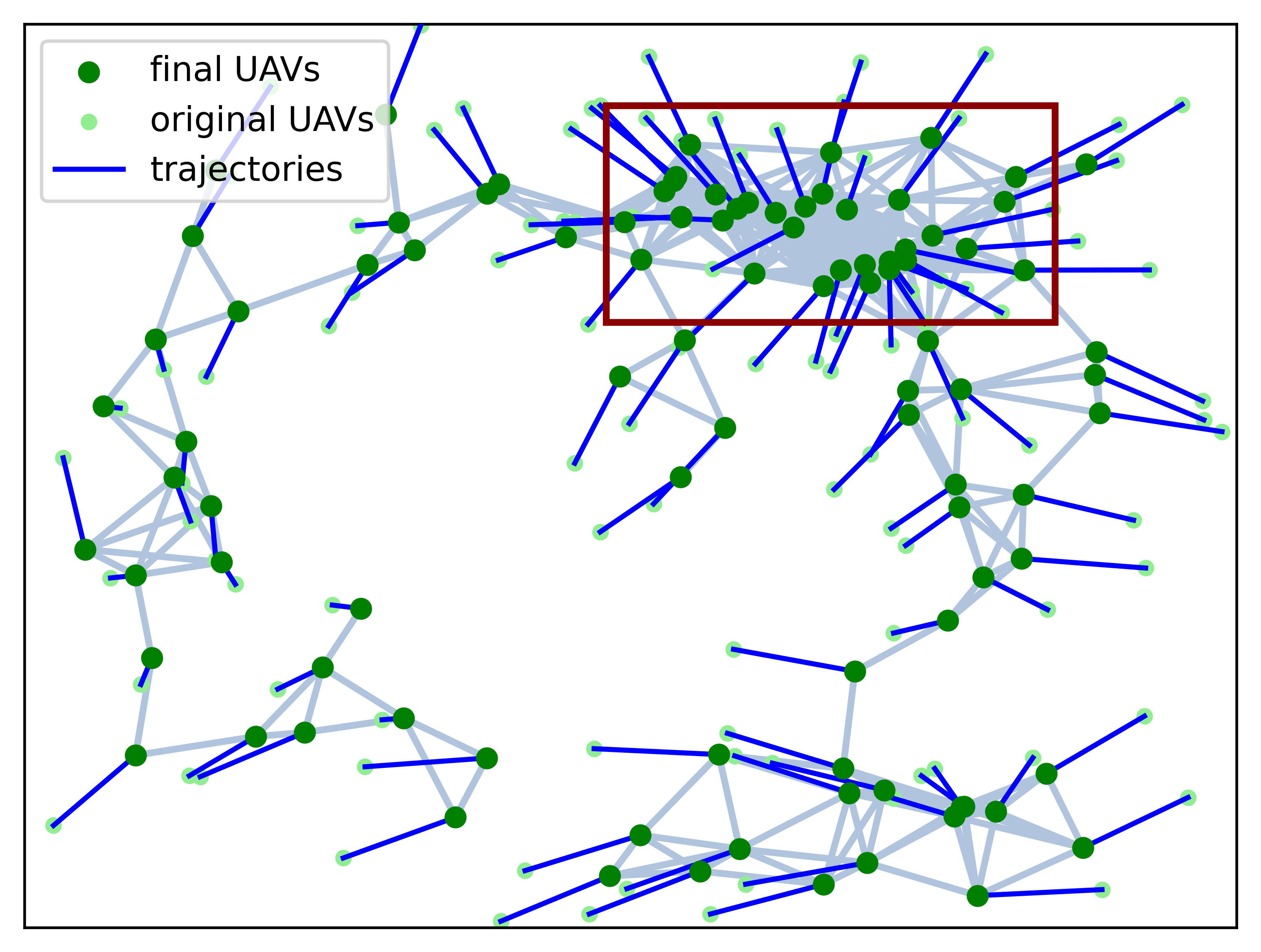}}}
\captionsetup{justification=raggedright, singlelinecheck=false}
\caption{An example of CNS issue caused by massive damage in USNET and recovery solutions of different algorithms.}
\vspace{-0.5em}
\label{cha}
\end{figure*}

\section{Related Work}
Resilient recovery algorithms for USNET have become a critical area of research, particularly in scenarios where connectivity is disrupted due to massive node destructions in hostile environments \cite{survey}. An example of CNS issue was illustrated in Fig.\ref{cha}a, where the USNET was divided into 7 sub-nets. Various strategies have been developed to address the CNS challenges, focusing on reconnecting divided sub-nets within minimum recovery time. 
Considering the different motivations for aggregating the surviving nodes, we can divide the existing approaches into the \textit{critical-node-based}, the \textit{meeting-point-based}, and the \textit{optimization-based} algorithms.

The critical-node-based algorithms aim to rebuild connectivity by relocating the remaining nodes to the locations of destroyed critical nodes. For example, algorithms in \cite{mv_cut1, mv_cut2} identified the global cut-vertex nodes as the critical nodes and addressed the CNS issues in small networks. Zhang \textit{et al.}\cite{mv_cut3} extended the recovery algorithm to large-scale WSNs by employing a finite state machine, effectively reducing the complexity of identifying cut-vertex nodes. Younis \textit{et al.}\cite{mv_ms1} considered a clustered USNET and designated the master nodes as the critical nodes, eliminating the need for additional identification processes. However, the cascaded motion under massive damage increases the total number and moving distances of the relocated nodes and will reduce the network's coverage ability \cite{mv_ms2}. As a result, the critical-node-based methods have poor performance in massive damage scenarios of swarm networks.

The meeting-point-based algorithms guide the remaining nodes to aggregate toward a pre-defined meeting point, which can ensure the restoration of connectivity. These algorithms have been widely applied in mobile WSNs \cite{mv_wsn1, mv_wsn2, mv_wsn3}, with the meeting point usually located at the network center. However, in large-scale USNETs where nodes are distributed across vast spaces, the edge nodes may be situated far from the meeting point. As a result, the recovery time increases because the edge nodes, shown in the red boxes in Fig.\ref{cha}b, must cover long distances to reach the meeting point. To cope with this problem, Mi \textit{et al.}\cite{hero} proposed to define multiple meeting points, which reduces the average moving distance of edge nodes. Chen \textit{et al.}\cite{sidr} developed a swarm intelligence-based damage-resilient mechanism that calculates the recovery path for each sub-network as a whole. This approach allows the network to restore connectivity without losing its original links and achieves path tracking with low communication overhead.

The optimization-based algorithms formulate the recovery issues as an optimization problem to find the optimal recovery paths for the remaining nodes \cite{mv_opt1, mv_opt2, mv_opt3}. With the development of UAV intelligence, more advanced approaches integrate deep learning and graph-based algorithms to evolve the network topologies. Mou \textit{et al.}\cite{mgc} applied the graph convolution operation (GCO) to solve the CNS challenge for the first time, where the information aggregation capability of GCO enables the algorithm to achieve state-of-the-art performance. However, due to the nature of graph convolution, which involves information aggregation between neighbors, nodes with higher degrees tend to dominate the convolution operation, attracting the remaining nodes towards them. This results in a serious over-aggregation in space, as shown in Fig.\ref{cha}c.

The optimization problem formulated by the CNS issue is generally non-convex, meaning that a well-chosen starting point for iteration can lead to a better solution. Inspired by the critical-node-based algorithms, we notice that the original USNET is connected and the destructed nodes can help direct the recovery paths towards the gaps between sub-nets. In our previous work \cite{demd}, we developed a damage-embedding module that maps the location information of destroyed nodes into the features of their remaining neighbors, which enhanced the performance of the GCO-based algorithm. This paper aims to explore an efficient model for characterizing, extracting, and leveraging network damage information, thereby achieving fast connectivity restoration and a more robust recovered topology.
%

\section{System model}
This section first constructs a graph topology to represent the UAV swarm network (USNET) and then proposes the massive damage model. The problem investigated in this paper is formulated in the third subsection.

\subsection{Graph Topology of USNET}
The original USNET is composed of $N$ massive independent and identical UAVs with a pre-defined connected structure as shown in Fig.\ref{sys}a, where the UAVs are labeled by a global index $\mathcal{I}=\{1,2,...,N\}$. Generally, UAVs form a network at equal altitudes for most missions, hence we mainly consider two-dimensional locations. Let location vector $\bm{p}_i(t)={[x_i(t),y_i(t)]}^\top \in \mathbb{R}^2$ denote the coordinates of the $i$-th UAV at time $t$, where $x_i(t)$ and $y_i(t)$ denote the coordinate component of $X$ and $Y$ axis, respectively.
Denote $\bm{X}(t)\in\mathbb{R}^{N\times 2}$ as the position matrix consisting of the location vectors of each UAV at time $t$. That is,
\begin{align}
    \bm{X}(t)=[\bm{p}_1(t),\bm{p}_2(t),...,\bm{p}_N(t)]^\top.
\end{align}

Let graph $G(t)=(\mathcal{U}(t),\mathcal{E}(t))$ denote the topology of the USNET, where $\mathcal{U}(t)=\{u_i|i\in \mathcal{I}(t)\}$ represents the node set and $\mathcal{E}(t)=\{e_{ij}|u_i,u_j\in \mathcal{U}(t)\}$ represents the edge set of communication links. Define the Eulerian distance between the UAV $u_i$ and $u_j$ at time $t$ as $d_{ij}(t)=\|\bm{p}_i(t)-\bm{p}_j(t)\|$, and $u_i$ and $u_j$ can establish a communication link when the distance is shorter than the UAV transmission range $d_{tr}$, i.e., the edge $e_{ij}\in \mathcal{E}(t)$ exists if and only if $d_{ij}(t)\leq d_{tr}$.

\begin{definition}[1-hop Neighboring Set]
Let $\mathcal{N}^1_i(t)$ be the set of neighbors of UAV $u_i$ at time $t$, i.e.,
\begin{align}
    \mathcal{N}^1_i(t)=\{u_j|e_{ij}\in \mathcal{E}(t)\}.
\end{align}
\end{definition}

\begin{definition}[$k$-hop Neighboring Set]
Let $\mathcal{N}^k_i(t)$ be the set of $k$th-hop neighbors of UAV $u_i$ at time $t$, e.g., node $u_j$ in Fig.\ref{sys}a is a 3-hop neighbor of $u_i$. In particular, we define ${N}^0_i(t)=\{u_i\}$ as the set of node $u_i$ itself. 
\end{definition}

Based on the $k$-hop neighboring sets, we then determine a multi-hop sub-graph of node $u_i$. Let graph $G^k_i(t)=(\mathcal{U}^k_i(t),\mathcal{E}^k_i(t))$ denote the local multi-hop topology of UAV $u_i$ at time $t$, where $\mathcal{U}^k_i(t)=\bigcup_{m=0}^k\mathcal{N}^{m}_i(t)$ contains all neighbors of $u_i$ within $k$-hop range and $\mathcal{E}^k_i(t)=\{e^k_{jl}|u_j,u_l\in \mathcal{U}^k_i(t)\}$ represents the link of multi-hop relations. The edge $e^k_{jl}$ exists if and only if $u_j\in \mathcal{U}^k_l(t)$, i.e., the hop number between $u_j$ and $u_l$ is shorter than $k$-hop.

The multi-hop sub-graph contains part of the nodes in $G(t)$, but the neighboring topology is enhanced to characterize the local relationships near each UAV. This method of reconstructing topology based on multi-hop relations is also known as \textit{graph diffusion}\cite{diffusion}.

\subsection{Damage Model and MDSG}
\begin{figure}[!t]
\vspace{-1em}
\centering{
\subfloat[original USNET]{\includegraphics[width=.5\linewidth]{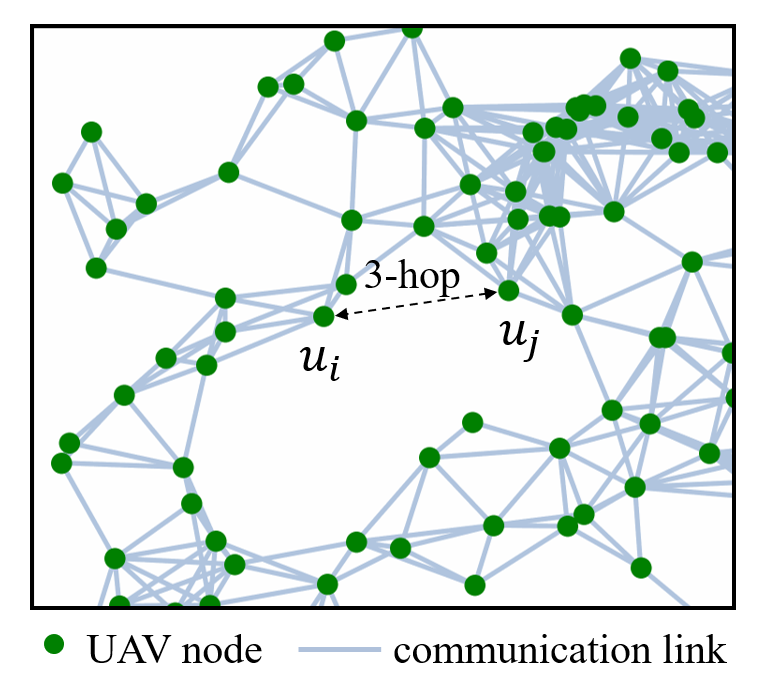}}
\subfloat[damaged USNET]{\includegraphics[width=.5\linewidth]{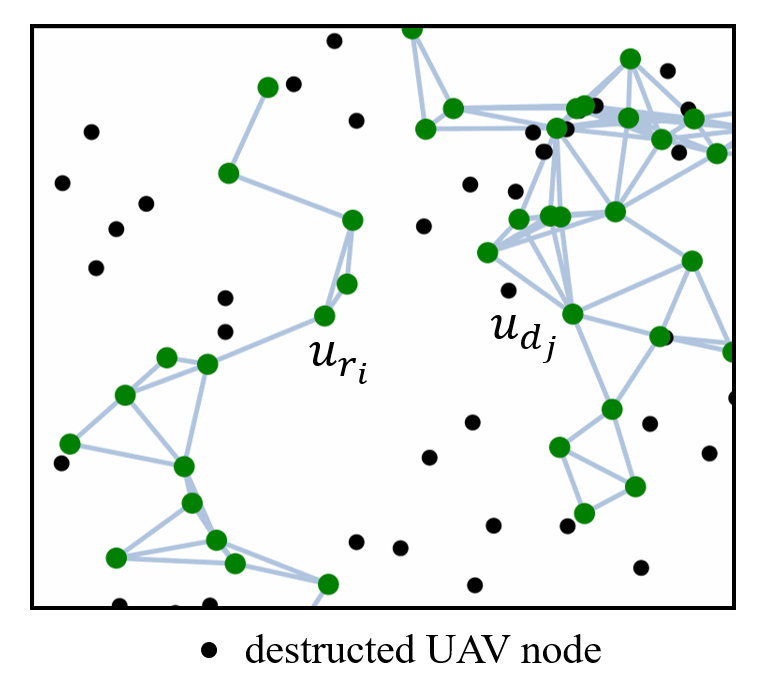}}\vspace{-1em}\\
\subfloat[network split]{\includegraphics[width=.5\linewidth]{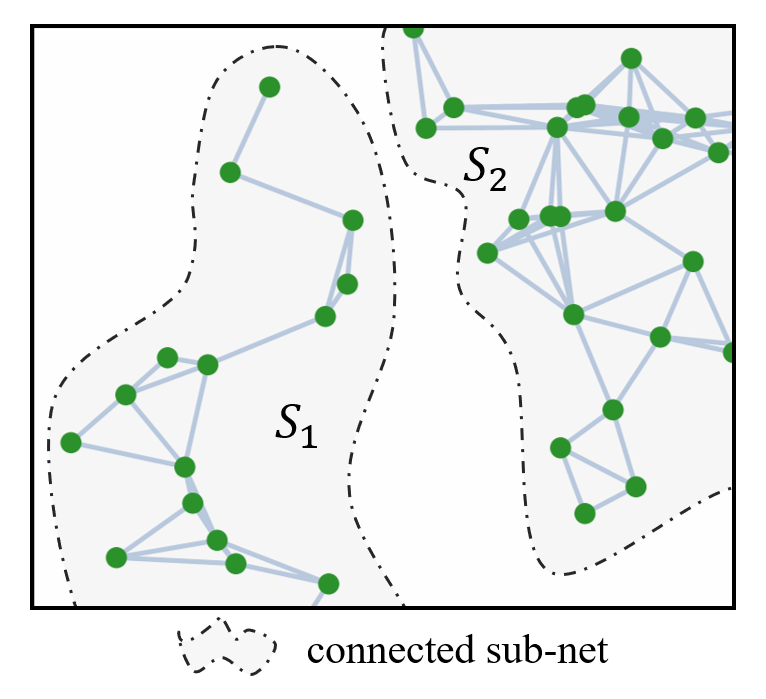}}
\subfloat[MDSG]{\includegraphics[width=.5\linewidth]{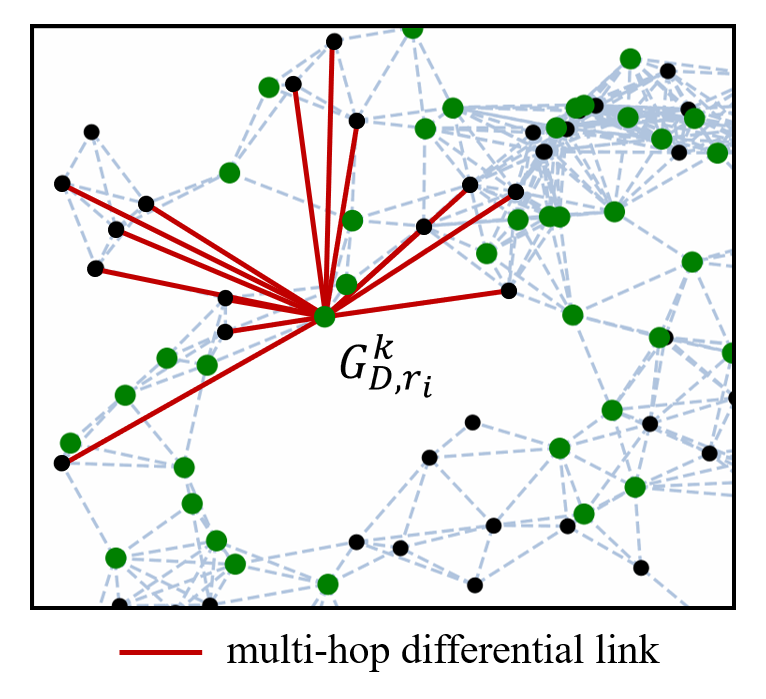}}}
\caption{An example of USNET graph topology and massive damage that causes split the network.}
\label{sys}
\end{figure}

In this paper, we focused on network-splitting scenarios caused by massive node destructions, which can occur simultaneously or cumulatively. As shown in Fig.\ref{sys}b, 50\% of UAV nodes were damaged, resulting in the \textit{communication network split} (CNS) issue. Therefore, we can determine the moment $t_0$ that the USNET happens to split with total $N_D$ UAVs were destroyed and $N_R=N-N_D$ UAVs remained. The damage index $\mathcal{I}_D=\{d_1,d_2,...,d_{N_D}\}$ labeled all destructed UAVs, and the remain index $\mathcal{I}_R=\{r_i|r_i\in \mathcal{I}-\mathcal{I}_D\}$ contained the remaining UAV nodes.

To model the topology after the damage, we constructed a \textit{Remained Graph} $G_R(t)=(\mathcal{U}_R(t),\mathcal{E}_R(t))$ at time $t\geq t_0$, where $\mathcal{U}_R(t)=\{u_{r_i}|r_i\in \mathcal{I}_R\}$ is the remaining-node set, and $\mathcal{E}_R(t)=\{e_{r_ir_j}|e_{r_ir_j}\in\mathcal{E}(t), r_i,r_j\in \mathcal{I}_R\}$ is the edge set for remaining UAVs. Due to the CNS issue, the \textit{Remained Graph} is divided into several disconnected sub-nets, as shown in Fig.\ref{sys}c. 

\begin{definition}[Disconnected Sub-nets]
Denote $\mathcal{S}_{all}(t)=\{\mathcal{S}_j(t)|\mathcal{S}_j(t)\subseteq\mathcal{U}_R(t)\}$ as the series of sub-nets in $G_R(t)$ at time $t$. Two set $\mathcal{S}_i(t)$ and $\mathcal{S}_j(t)$ are said to be disconnected at time $t$ if $\forall u_k\in \mathcal{S}_i(t)$ can not reach other nodes $\forall u_l\in \mathcal{S}_j(t)$.
\end{definition}

\begin{definition}[Number of Sub-nets]
Denote $N_s(t)\in \mathbb{N}$ as the number of sub-nets in $G_R(t)$ at time $t$, i.e., $N_s(t)=|\mathcal{S}_{all}(t)|$. Apparently, $G_R(t)$ is connected at time $t$ if and only if $N_s(t)=1$. 
\end{definition}

A widely-used method to calculate $N_s(t)$ of $G_R(t)$ with the Laplace matrix is introduced in \cite{laplacian}. We first define the adjacency matrix of $G(t)$ at time $t$ as $\bm{A}(t)=(a_{ij})\in \mathbb{S}^{|\mathcal{I}(t)|}$, where $a_{ij}$ is a 0-1 variable to denote if $e_{ij}$ exist or not, and $\mathbb{S}^{|\mathcal{I}(t)|}$ represents the set of symmetric matrices. Similarly, the adjacency matrix of the \textit{Remained Graph} $G_R(t)$ at time $t$ is defined as $\bm{A}_R(t)=(a_{r_ir_j})$, and $a_{r_ir_j}=1$ if and only if $e_{r_ir_j}\in \mathcal{E}_R(t)$. With the adjacency matrix $\bm{A}_R(t)$, we have the degree matrix of $G_R(t)$ as
\begin{equation}
\bm{D}_R(t)={\rm diag}(d_{r_1}(t),d_{r}(t),...,d_{r_{N_R}}(t))
\label{degree}
\end{equation}
where $d_{r_i}(t)=\sum_{j=r_1}^{r_{N_R}}a_{r_ir_j}(t)$ is the degree of each remaining node, and ${\rm diag}(\cdot)$ represents a diagonal matrix. The Laplace matrix of $G_R(t)$ is define as the difference between $\bm{D}_R(t)$ and $\bm{A}_R(t)$, i.e., 
\begin{equation}
    \bm{L}_R(t)=\bm{D}_R(t)-\bm{A}_R(t).
    \label{laplace}
\end{equation}

Since the $\bm{L}_R(t)$ is a positive semi-definite matrix, with the eigenvalue decomposition applied on $\bm{L}_R(t)$, we have
\begin{equation}
    \bm{L}_R(t)=\bm{U}_R(t)\bm{\Lambda}_R(t)\bm{U}_R^\top(t)
\end{equation}
where $\bm{U}_R(t)=[\bm{u}_{r_1}(t),\bm{u}_{r}(t),...,\bm{u}_{r_{N_R}}(t)]$ as the unitary matrix composed of orthogonal eigenvectors, and $\bm{\Lambda}_R(t)={\rm diag}(\lambda_{r_1}(t),\lambda_{r}(t),...,\lambda_{r_{N_R}}(t))$ denotes the matrix with non-negative eigenvalues. The number of sub-net is equal to the number of zero eigenvalues in $\bm{\Lambda}_R(t)$, i.e., 
\begin{align}
    N_s(t)={\rm Count}(\lambda=0|\bm{\Lambda}_R(t)).
\end{align}

The \textit{Remained Graph} characterized the severity of network split, and the differential topology before and after the CNS provided valuable information for network recovery. This information was carried on the destructed nodes $\mathcal{U}_D(t)=\{u_{d_i}|d_i\in \mathcal{I}_D\}$ and the damaged links, hence we proposed the \textit{multi-hop differential sub-graph} (MDSG) to represent the local damage for each remaining nodes. 

\begin{definition}[MDSG]
Let a static graph $G^k_{d,r_i}(t_0)=(\mathcal{U}^k_{d,r_i},\mathcal{E}^k_{d,r_i})$ denote the multi-hop differential topology of remaining UAV $u_{r_i}$ at time $t_0$, where $\mathcal{U}^k_{d,r_i}=\mathcal{U}^k_{r_i}\bigcap\mathcal{U}_D(t_0)$ contains all destructed neighboring nodes of $u_{r_i}$ within $k$-hop range, and $\mathcal{E}^k_{d,r_i}(t)=\{e^k_{d,r_id_j}|u_{d_j}\in \mathcal{U}^k_{d,r_i}\}$ represents the links between $u_{r_i}$ and its destructed neighbors. The edge $e^k_{d,r_id_j}$ exists for every $u_{d_j}\in \mathcal{U}^k_{d,r_i}$.
\end{definition}

The 3-hop differential sub-graph of given $u_{r_i}$ was illustrated in Fig.\ref{sys}d. The MDSG characterized the multi-hop relationship between the remaining nodes and the damaged nodes in the original network topology, but there was no link between the remaining node pairs or destructed node pairs, hence the MDSG was a bipartite graph.

\subsection{Problem Formulation}
The purpose of CNS recovery is to reform connectivity of \textit{Remained Graph} $G_R(t)$ with minimum recovery time, which is achieved by the movement of the UAVs. Let time $t_0<t_1<t<...<t_n$ denote a discrete time series, where the time step $\Delta t=t_{m+1}-t_m$ is constant. Since network recovery is a continuous process, the recovery methods need to calculate the speed and direction of each moment $t_m$ for the UAV nodes.

Let $\bm{V}_R(t_m)=[\bm{v}_{r_1}(t_m),\bm{v}_{r}(t_m),...,\bm{v}_{r_{N_R}}(t_m)]^\top$ denote the matrix of velocity vectors of the remaining nodes at time $t_m$, where $\bm{v}_{r_i}(t_m)\in \mathbb{R}^2$. The positions of each remaining node at time $t_m$ are then calculated as
\begin{equation}
\begin{aligned}
    \bm{p}_{r_i}(t_m)=&\bm{p}_{r_i}(t_{m-1})+\Delta t \cdot\bm{v}_{r_i}(t_{m-1})\\
    =&\bm{p}_{r_i}(t_0)+\mathop{\sum}\limits_{t_0\leq t_l\leq t_m}\Delta t \cdot\bm{v}_{r_i}(t_l)
\end{aligned}
\end{equation}
and the new position matrix of the \textit{Remained Graph} $\bm{X}_R(t_m)=[\bm{p}_{r_1}(t_m),\bm{p}_{r}(t_m),...,\bm{p}_{r_{N_R}(t_m)}]^\top$ updates the adjacent matrix $\bm{A}_R(t_m)$ and the number of sub-nets $N_S(t_m)$. 

For each time step $t_m$ in the recovery process, the goal of UAV movement is to minimize the number of sub-nets in the \textit{Remained Graph} for the next time step, i.e.,
\begin{align}
    \mathop{\rm min}\limits_{\bm{V}_R(t_m)}\quad & N_S(t_{m+1})\\
    {\rm s.t.}\quad & \|\bm{v}_{r_i}(t_m)\|\leq v_{max}, r_i\in \mathcal{I}_R
\end{align}
where $v_{max}$ is the maximum flight speed of UAVs.

Note that the number of sub-nets has a lower boundary $inf\{N_S(t)\}=1$ that indicates the \textit{Remained Graph} has already connected. Since (P1) is always decreasing the value of $N_S(t)$, we can determine the time $t_r$ when $N_S(t)$ reaches 1, and the recovery time cost is $T_{rc}=t_r-t_0$. The goal of this paper is to find a series of velocity matrix $\mathcal{V}_R=\{\bm{V}_R(t_l)|t_0\leq t_l<t_r\}$ that minimizes the recovery time, i.e.,
\begin{align}
    ({\rm P}1):\quad \mathop{\rm min}\limits_{\mathcal{V}_R,t_r}\quad & T_{rc}=t_r-t_0\\
    \quad {\rm s.t.}\quad & N_S(t_r) = 1\\
    & \|\bm{v}_{r_i}(t_l)\|\leq v_{max}, t_0\leq t_l<t_r, r_i\in \mathcal{I}_R.
    \label{consv}
\end{align}

\textit{Notations}: $x,\bm{x},\bm{X},\mathcal{X}$ represent a scalar $x$, a vector $\bm{x}$, a matrix $\bm{X}$, and a set $\mathcal{X}$, respectively; $\partial, \propto, \ast$ denote partial derivative operator, proportional operator, and convolution operator, respectively; $\top$ is the transpose of vectors and matrices; $\bigcup, \bigcap$ denote the union the intersection operator for sets, respectively; $\mathbb{R},\mathbb{N},\mathbb{S}$ represent the real number space, the natural number space, and the symmetric matrix space;
$|\mathcal{X}|$ represents the number of elements in set $\mathcal{X}$; $\|\bm{x}\|$ represents the norm of vector $\bm{x}$; $\|\bm{X}\|_\infty$ is the infinite norm of matrix $\bm{X}$; $\lfloor\cdot\rfloor$ denotes the floor function;
$inf\{\cdot\}$ and $sup\{\cdot\}$ are the lower bound and upper bound of sets, respectively.

\section{MDSG-based Artificial Potential Field Algorithm}
The artificial potential field (APF) algorithm is one of the classic methods for UAV path planning \cite{APF}, which calculates the velocity or acceleration for the UAVs by simulating the virtual force field to reach the destination location. Therefore, the performance of an APF algorithm is mainly affected by the design of the force field. Since the remaining UAVs are supposed to fill the gaps between sub-nets, the differential topology can guide the remaining UAVs toward the locations of destructed UAVs. In this section, we proposed a \textit{Multi-hop Differential Sub-Graph based Artificial Potential Field} (MDSG-APF) algorithm, aiming to utilize the damage information in a simple but efficient way for CNS recovery.

The section includes the construction of the force field based on MDSG, the APF-based solution for (P1), and the algorithm analysis on convergence and recovery time, as will be stated as follows.


\subsection{MDSG-based Force Field}

\begin{figure}[!t]
\vspace{-1em}
\centering{
\subfloat[convergent force field]{\includegraphics[width=.5\linewidth]{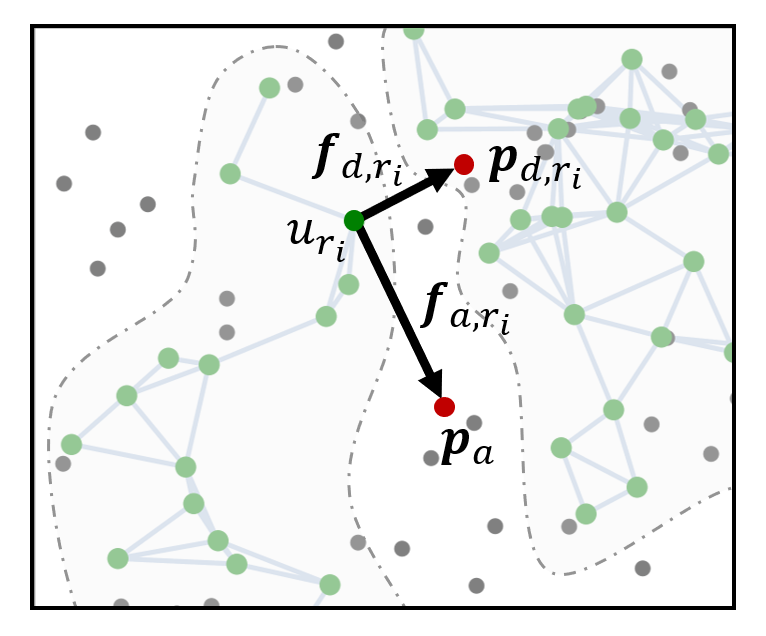}}
\subfloat[divergent force field]{\includegraphics[width=.5\linewidth]{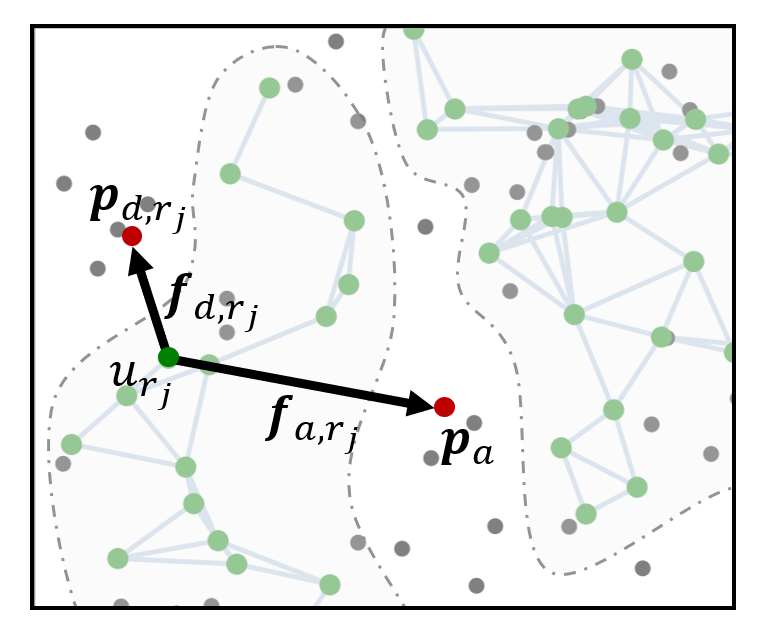}}}
\captionsetup{justification=raggedright, singlelinecheck=false}
\caption{Examples of MDSG-based force fields for remaining UAV nodes with different locations.}
\label{field}
\end{figure}

The MDSG $G_{d,r_i}^k$ of node $u_{r_i}$ contains the location information about its destructed $k$-hop neighbor UAVs, and each destructed neighbor $u_{d_j}\in \mathcal{U}^k_{D,r_i}$ provides a node-level gravitational force $\bm{f}_{d,r_id_j}$ for $u_{r_i}$. This node-level force is directed by location $\bm{p}_{r_i}(t_0)$ to $\bm{p}_{d_j}(t_0)$, hence the resultant force $\bm{f}_{d,r_i}$ composed of all destructed neighbors is summed as $\bm{f}_{d,r_i}=\sum\bm{f}_{d,r_id_j}$. In other word, let position vector $\bm{p}_{d,r_i}$ denote the central location of MDSG $G^k_{d,r_i}$ as
\begin{align}
    \bm{p}_{d,r_i} = \frac{1}{|\mathcal{U}^k_{D,r_i}|}\cdot\mathop{\sum}\limits_{u_{d_j}\in\mathcal{U}^k_{D,r_i}}\bm{p}_{d_j}(t_0),
    \label{pd}
\end{align}
we have the resultant force $\bm{f}_{d,r_i}$ given by MDSG is directed from location $\bm{p}_{r_i}(t_0)$ to location $\bm{p}_{d,r_i}$, so called the \textit{differential force}. When $u_{r_i}$ locates near the boundary of its sub-net, the destructed neighbors in MDSG are mainly distributed in the gaps between sub-nets, hence the force $\bm{f}_{d,r_i}$ points at the gap nearby as shown in Fig.\ref{field}a. 

However, when the remaining UAV is located on the other side of its sub-net, e.g., node $u_{r_j}$ in Fig.\ref{field}b, the force $\bm{f}_{d,r_j}$ can point to a direction away from the USNET, leading to a divergent recovery method. To guarantee the final connectivity of the \textit{Remained Graph}, we form an \textit{aggregation force} $\bm{f}_{a,r_i}$ for each remaining node to balance the direction of $\bm{p}_{d,r_i}$. Let position vector $\bm{p}_{a}$ denote the global central location of all destructed nodes, that is,
\begin{align}
    \bm{p}_{a} = \frac{1}{N_D}\cdot\mathop{\sum}\limits_{d_j\in\mathcal{I}_D}\bm{p}_{d_j}(t_0).
    \label{pa}
\end{align}

The \textit{aggregation force} has a direction from location $\bm{p}_{r_i}(t_0)$ to the global aggregation location $\bm{p}_{a}$. Therefore, the force field of each remaining UAV node is composed of a \textit{differential force} that represents the direction of damage gaps nearby, and an \textit{aggregation force} to guarantee the direction of aggregation.

\subsection{APF Algorithm Solution}
With the force field calculated according to MDSG, we can form the artificial potential field for the remaining UAVs to generate the recovery directions.
Let function $\phi_{r_i}(t)$ denote the potential field of remaining node $u_{r_i}$ at time $t$, we have the velocity of $u_{r_i}$ calculated as
\begin{align}
    \bm{v}_{r_i}(t) = \bm{p}'_{r_i}(t) = -\frac{\partial\phi_{r_i}(t)}{\partial\bm{p}_{r_i}(t)}.
    \label{vri}
\end{align}

According to Newton's second law, the force is positively correlated with the acceleration, i.e., the derivative of the node velocity, that is,
\begin{align}
    \bm{f}_{r_i}(t)\propto \bm{v}'_{r_i}(t)= -\frac{\partial\phi_{r_i}(t)}{\partial\bm{p}_{r_i}(t)\cdot \partial t}.
\end{align}
where $\bm{f}_{r_i}(t)=\bm{f}_{d,r_i}+\bm{f}_{a,r_i}$ is the resultant force of MDSG-based force field of $u_{r_i}$ at time $t$. Since the locations of destructed nodes were no longer updated after the massive damage, the forces $\bm{f}_{d,r_i}$ and $\bm{f}_{a,r_i}$ only exist at the time moment $t_0$. When the remaining UAV had left the location $\bm{p}_{r_i}(t_0)$, the two forces provided by MDSG at time $t_0$ become meaningless and can be regarded as 0. Therefore, the accelerations of remaining UAVs are equal to 0 at $t>t_0$, which means the velocity of each remaining node is constant in the following recovery process, i.e, $\bm{v}_{r_i}(t_0)=\bm{v}_{r_i}(t_1)=...=\bm{v}_{r_i}(t_r)$, where $t_0<t_1<...<t_r$ is the time series of recovery. In conclusion, the APF algorithm only needs to calculate the initial recovery velocity $\bm{v}_{r_i}(t_0)$ for each remaining node $u_{r_i}$. 

The potential function is supposed to generate two forces for each remaining node at time $t_0$: the \textit{differential force} caused by filling local network gaps, and the \textit{aggregation force} to guarantee the connectivity. Therefore, the potential function is also composed of two parts as
\begin{align}
    \phi(t_0) = \mathop{\sum}\limits_{r_i\in \mathcal{I}_R}\phi_{r_i}(t_0)=\mathop{\sum}\limits_{r_i\in \mathcal{I}_R}(\alpha_{r_i}\phi_{d,r_i}(t_0)+\phi_{a,r_i}(t_0))
    \label{devide}
\end{align}
where $\phi_{d,r_i}(t_0)\in \mathbb{R}$ and $\phi_{a,r_i}(t_0)\in \mathbb{R}$ denote the potential functions of the \textit{differential force} and the \textit{aggregation force} of $u_{r_i}$ at time $t_0$, respectively, and $\alpha_{r_i}\geq 0$ is the weight to balance the forces.
The potential functions of two forces based on MDSG are defined as
\begin{align}
    \phi_{d,r_i}(t_0) = \frac{1}{2}|\bm{p}_{d,r_i}-\bm{p}_{r_i}(t_0)|^\top &|\bm{p}_{d,r_i}-\bm{p}_{r_i}(t_0)|\\
    \phi_{a,r_i}(t_0) = \frac{1}{2}|\bm{p}_a-\bm{p}_{r_i}(t_0)|^\top &|\bm{p}_a-\bm{p}_{r_i}(t_0)|
\end{align}
where $\bm{p}_{d,r_i}$ is the central location of MDSG given by (\ref{pd}) and $\bm{p}_a$ is the global central location given by (\ref{pa}). 

\begin{figure}[!t]
\centerline{\includegraphics[width=.75\linewidth]{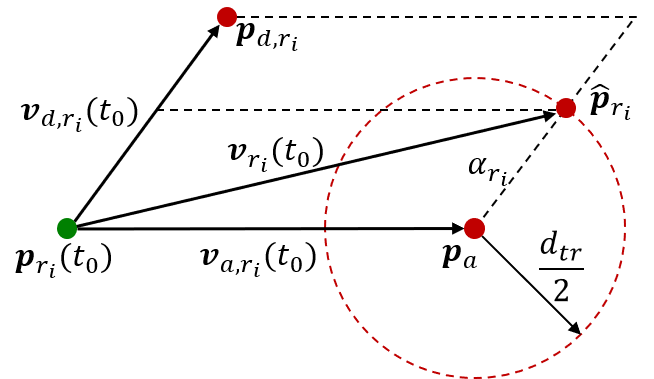}}
\captionsetup{justification=raggedright, singlelinecheck=false}
\caption{Velocity components of node $u_{r_i}$.}
\label{velocity}
\end{figure}

According to (\ref{vri}) and (\ref{devide}), the initial velocity of $u_{r_i}$ was decomposed into the differential velocity component $\bm{v}_{d,r_i}(t_0)$ and the aggregation velocity component $\bm{v}_{d,r_i}(t_0)$, as shown in Fig.\ref{velocity}. Hence, we have 
\begin{align}
    \bm{v}_{r_i}(t_0)=\alpha_{r_i}\bm{v}_{d,r_i}(t_0)+\bm{v}_{a,r_i}(t_0)
\end{align}
where the two components are calculated as 
\begin{align}
    \bm{v}_{d,r_i}(t_0) &= -\frac{\partial\phi_{d,r_i}(t_0)}{\partial\bm{p}_{r_i}(t_0)} = \bm{p}_{d,r_i}-\bm{p}_{r_i}(t_0)
    \label{vd}\\
    \bm{v}_{a,r_i}(t_0) &= -\frac{\partial\phi_{a,r_i}(t_0)}{\partial\bm{p}_{r_i}(t_0)} = \bm{p}_{a}-\bm{p}_{r_i}(t_0).
    \label{va}
\end{align}

To ensure the final graph is connected, the weight $\alpha_{r_i}$ was selected to balance the two velocity components so that $u_{r_i}$ won't move away from other sub-nets. The basic idea is to limit the final location $\hat{\bm{p}}_{r_i}$ of the remaining UAV $u_{r_i}$ within a circle area that the diameter is equal to the wireless transmission distance $d_{tr}$. This circle needs to be notified to each remaining node as global information, hence we set the location $\bm{p}_a$ as the center of the circle. According to Fig.\ref{velocity}, the weight $\alpha_{r_i}$ is calculated as
\begin{align}
    \alpha_{r_i} = \frac{d_{tr}}{2}\cdot\frac{1}{\|\bm{v}_{d,r_i}(t_0)\|} = \frac{d_{tr}}{2\|\bm{p}_{d,r_i}-\bm{p}_{r_i}(t_0)\|}
\end{align}

The resultant velocity is represented as
\begin{align}
    \bm{v}_{r_i}(t_0) = \frac{d_{tr}\cdot(\bm{p}_{d,r_i}-\bm{p}_{r_i}(t_0))}{2\|\bm{p}_{d,r_i}-\bm{p}_{r_i}(t_0)\|} + \bm{p}_{a}-\bm{p}_{r_i}(t_0)
    \label{v}
\end{align}

Since the UAV speed has a maximum value $v_{max}$, the final velocity $\hat{\bm{v}}_{r_i}(t_0)$ needs to be contracted to a proper value, that is,
\begin{align}
    \hat{\bm{v}}_{r_i}(t_0) = \frac{v_{max}}{\mathop{\rm max}\limits_{r_j\in \mathcal{I}_R}\|\bm{v}_{r_j}(t_0)\|}\cdot\bm{v}_{r_i}(t_0)
    \label{vhat}
\end{align}
where $\mathop{\rm max}\limits_{r_j\in \mathcal{I}_R}\|\bm{v}_{r_j}(t_0)\|$ is the maximum velocity among all remaining nodes given by (\ref{v}). The contraction of velocity limits the remaining nodes arriving at the circle area simultaneously, hence guaranteeing the restoration of connectivity according to Proposition \ref{prop-c} in the subsequent section.

Let $\bm{\hat{V}}_R(t_0)=[\bm{\hat{v}}_{r_1}(t_0),\bm{\hat{v}}_{r}(t_0),...,\bm{\hat{v}}_{r_{N_R}}(t_0)]^\top$ denote the initial velocity matrix after scaling, we have the APF algorithm solution for (P1) as
\begin{align}
    \hat{\mathcal{V}}_R=\{\hat{\bm{V}}_R(t_l)|t_0\leq t_l<t_r\}
    \label{apfv}
\end{align}
where $\hat{\bm{V}}_R(t_l)=\hat{\bm{V}}_R(t_0)$ is constant for the entire recovery process. It is easy to know that the maximum velocity constraint (\ref{consv}) is satisfied as $\|\hat{\bm{v}}_{r_i}(t_0)\|=\|\hat{\bm{v}}_{r_i}(t_0)\|\leq v_{max}$.

The network recovery process by the MDSG-APF algorithm solution is shown in Alg.\ref{alg1}. The process started at time $t_0$ when the massive damage occurred, and the number of sub-nets was determined as the judgment of CNS issue. Steps 2-4 calculated the velocity solution provided by the MDSG-APF algorithm for network recovery. Step 5 described the termination condition of the recovery process, that is, only one sub-net exists in the network. Steps 6-9 updated the velocity matrix, the position matrix, and the current time at each time step, then calculated the number of sub-nets in the new network graph. Finally, step 11 returned the final recovery time as output when the network connectivity was reformed.

\begin{algorithm}[!t]
    \caption{Network recovery process by the MDSG-APF algorithm solution.}
    \label{alg1}
    \begin{algorithmic}[1]
    \renewcommand{\algorithmicrequire}{\textbf{Input:}}
    \REQUIRE
    Initial time $t_0$, indexes $\mathcal{I}_R$ and $\mathcal{I}_D$, MDSG $G^k_{d,r_i}(t_0)$ for $u_{r_i}$, remained position matrix $\bm{X}_R(t_0)$
    \renewcommand{\algorithmicensure}{\textbf{Output:}}
    \ENSURE
    A covered USNET, final recovery time $t_r$.
    \STATE Initialize time $t_l=t_0$, determine $N_S(t_0)$
    \STATE Calculate $\bm{p}_{d,r_i}$ and $\bm{p}_a$ based on $G^k_{d,r_i}(t_0)$
    \STATE Calculate $\alpha_{r_i}$ and $\bm{v}_{r_i}(t_0)$
    \STATE Calculate solution $\hat{\bm{v}}_{r_i}(t_0)$ and $\hat{\bm{V}}_R(t_0)$
    \WHILE{$N_S(t_l)>1$}
    \STATE $\hat{\bm{V}}_R(t_l)\leftarrow\hat{\bm{V}}_R(t_0)$
    \STATE $\bm{X}_R(t_{l+1})\leftarrow\bm{X}_R(t_l)+\Delta t\cdot \hat{\bm{V}}_R(t_l)$
    \STATE Update $l=l+1$
    \STATE Determine new $N_S(t_l)$
    \ENDWHILE
    \RETURN $t_r=t_l$
    \end{algorithmic}
\end{algorithm}

\subsection{Theoretical Analysis of MDSG-APF}
The recovery method for the CNS issue aims to reform the \textit{Remained Graph} into a connected one. Let $\hat{G}_R(t_r)$ denote the final graph of USNET recovered by the APF solution $\hat{\mathcal{V}}_R$ in (\ref{apfv}), we then analyze how the solution meets the constraints in (P1), and discuss the theoretical upper bound of recovery time.

\subsubsection{Convergence of MDSG-APF}
The algorithm convergence requires the number of sub-nets $N_S(t_r)$ in the final graph $\hat{G}_R(t_r)$ equals 1. Since the final UAV nodes were located within a circle area, the connection between each node was reformed at time $t_r$. Proposition \ref{prop-c} proves the convergence of MDSG-APF.

\begin{proposition}
The \textit{Remained Graph} is connected if the remaining nodes are located within the circle area with a diameter equal to $d_{tr}$.
\label{prop-c}
\end{proposition}

\begin{figure}[!t]
\centerline{\includegraphics[width=.6\linewidth]{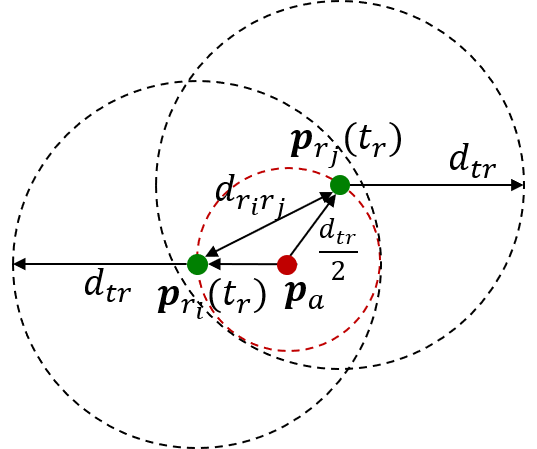}}
\captionsetup{justification=raggedright, singlelinecheck=false}
\caption{Connection of remaining nodes $u_{r_i}$ and $u_{r_j}$ within the circle.}
\label{convergence}
\end{figure}

\begin{IEEEproof}
Assume that the central location of the circle is $\bm{p}_a$.
For $\forall r_i,r_j\in \mathcal{I}_R$, if $u_{r_i}$ and $u_{r_j}$ are located within the circle at time $t_r$, the distances from locations $\hat{\bm{p}}_{r_i}$ and $\hat{\bm{p}}_{r_j}$ of each node to the center $\bm{p}_a$ of the circle are not larger than the radius, i.e., we have $\|\hat{\bm{p}}_{r_i}-\bm{p}_a\|\leq\frac{d_{tr}}{2}$ and $\|\hat{\bm{p}}_{r_j}-\bm{p}_a\|\leq\frac{d_{tr}}{2}$. Let $d_{r_ir_j}$ denote the distance between $u_{r_i}$ and $u_{r_j}$. As shown in Fig.\ref{convergence}, we have $d_{r_ir_j}\leq \|\hat{\bm{p}}_{r_i}-\bm{p}_a\|+\|\hat{\bm{p}}_{r_j}-\bm{p}_a\|\leq d_{tr}$, which indicates that $u_{r_i}$ can form a communication link with $u_{r_j}$. Therefore, the \textit{Remained Graph} becomes a completely connected graph with only one sub-net. This completes the proof.
\end{IEEEproof}

\subsubsection{Upper Bound of Recovery Time}
According to Proposition \ref{prop-c}, the connectivity of the \textit{Remained Graph} is reformed when all remaining nodes reach the circle at time $t_r$, and the total recovery time cost is calculated as $T_{rc}=t_r-t_0$. Since the velocity solution of the MDSG-APF algorithm is constant during the recovery process, each remaining UAV $u_{r_i}$ flies directly toward its final location $\hat{\bm{p}}_{r_i}$. The recovery time cost of the worst case is given in Proposition \ref{prop-t}, which is also the upper bound of recovery time by solution $\hat{\mathcal{V}}_R$.

\begin{proposition}
When massive damage occurs in the USNET, it takes at most $T^{max}_{rc}$ time to move all remaining nodes into a circle area and reform the connectivity of the \textit{Remained Graph}. The upper bound $T^{max}_{rc}$ is calculated as
\begin{align}
    T^{max}_{rc}=sup\{T_{rc}\}=\frac{\mathop{\rm max}\limits_{r_j\in \mathcal{I}_R}\|\bm{p}_a-\bm{p}_{r_j}(t_0)\|+\frac{d_{tr}}{2}}{v_{max}}.
\end{align}
\label{prop-t}
\end{proposition}

\begin{IEEEproof}
See Appendix A.
\end{IEEEproof}

In Proposition \ref{prop-t} we discussed the theoretical maximum recovery time under the worst case of massive damage scenario. Normally, the network connectivity is restored at an earlier time by the solution of the MDSG-APF algorithm. Compared with other meeting-point-based methods mentioned in Section \uppercase\expandafter{\romannumeral2}, the MDSG-APF allows each remaining node to move toward the local network gaps with maximum $\frac{d_{tr}}{2}$ of flight distance, which provides the effectiveness of the final recovery process. The MDSG-APF achieves the recovery of split networks with low complexity and thus is more suitable for the USNET with low intelligence and limited computing resources.

\section{MDSG-based Graph Convolution}
Let us assume that the UAVs in the USNET possess advanced intelligence, allowing for the application of more complex methods to address the CNS issue. Referring to the system model outlined in Section \uppercase\expandafter{\romannumeral3}, the USNET inherently forms a graph structure, with nodes representing UAVs and edges representing communication links. Given that the graph convolution operation (GCO) adaptively aggregates node features and updates relationship weights, it is well-suited to effectively capture the critical information from the damaged nodes in the MDSG.

In this section, we propose a \textit{Multi-hop Differential Sub-Graph based Graph Convolution} (MDSG-GC) algorithm designed to achieve faster recovery by analyzing the relationships between the damaged and intact nodes. The subsequent subsections first present the novel bipartite graph convolution tailored for the unique topology of the MDSG. We then introduce the structure of the MDSG-GC algorithm, which provides the solution to problem (P1).

\subsection{Design of Bipartite Graph Convolution}
The MDSGs capture the local neighboring relationships between the damaged UAVs and the remaining UAVs. However, addressing the CNS issue requires a coordinated effort from all remaining nodes. Since our recovery method aims to determine the trajectory based on the information from the damaged nodes, we constructed a unified multi-hop differential graph that integrates data from both the damaged and remaining UAVs, incorporating the relationships represented in the MDSGs.

\subsubsection{United MDSG}
The MDSG for each remaining node represents local links with limited scope but single MDSG can not provide a solution for all remaining nodes. To address this, we need to integrate all these MDSGs into a \textit{United Graph}, which encompasses the features of every UAV from the original USNET prior to the destruction, along with the multi-hop differential relationships on a global scale.
Let the united multi-hop differential graph $G_d^k=(\mathcal{U}_u,\mathcal{E}_{d,u}^k)$ with a fixed index $\mathcal{I}_u=\{\mathcal{I}_R,\mathcal{I}_D\}$ of all $N$ UAVs, where $k$ is the hop number, $i_a^v$ is the label of the virtual aggregation node, $\mathcal{U}_u=\{\mathcal{U}_R(t_0),\mathcal{U}_D(t_0)\}$ is resorted by $\mathcal{I}_u$, and $\mathcal{E}_{d,u}^k$ denotes the multi-hop neighboring relations according to the MDSGs. Likewise, the edge $e_{d,r_id_j}^k\in \mathcal{E}_{d,u}^k$ exists for every $u_{d_j}\in \mathcal{U}_{d,r_i}^k$ with the MDSGs of each $u_{r_i}$.
The position matrix of $G_d^k$ at time $t_0$ is represented as 
\begin{align}
    \bm{X}_{d}(t_0)=[\bm{p}_{r_1}(t_0),...,\bm{p}_{r_{N_R}}(t_0),\bm{p}_{d_1}(t_0),...,\bm{p}_{d_{N_D}}(t_0)]^\top
\end{align}
where $\bm{p}_{r_i}(t_0)$ and $\bm{p}_{d_j}(t_0)$ denote the locations of the UAV $u_{r_i}$ and $u_{d_j}$ at time $t_0$, respectively, $p_a$ is the aggregation location given by (\ref{pa}).

Define the adjacency matrix of the \textit{United Graph} $G_d^k$ as $\bm{A}_d^k=(a_{d,ij}^k)$, the value of each $a_{d,ij}^k$ relies on the edge set $\mathcal{E}_{d,u}^k$, that is, $a_{d,ij}^k$ equals 1 if $u_i$ and $u_j$ are $k$-hop differential neighbors in MDSG. Note that $\mathcal{E}_{d,u}^k$ contains only links between destructed nodes and remaining nodes, the \textit{United Graph} $G_d^k$ is a bipartite graph with the adjacency matrix as
\begin{align}
    \bm{A}_d^k = 
    \begin{pmatrix}
    \bm{0}_{N_R} & \bm{A}_{d,r_id_j}^k \\
    \bm{A}_{d,r_id_j}^{k\top} & \bm{0}_{N_D}
    \end{pmatrix}
\end{align}
where $\bm{0}_n\in \mathbb{R}^{n\times n}$ denotes the zero matrix, $\bm{A}_{d,r_id_j}^k$ denotes the $k$-hop differential links between $u_{r_i}$ and $u_{d_j}$. Due to the special topology of $G_d^k$, we improved a bipartite graph convolution operation for the $G_d^k$.

\subsubsection{Bipartite graph convolution operation}
We can get the Laplace matrix $\bm{L}_d^k$ of the \textit{United Graph} $G_d^k$ according to (\ref{degree}) and (\ref{laplace}). The features of $G_d^k$ form a structure in non-Euclidean space, hence the features need to be mapped from the time domain into the frequency domain to achieve the convolution operation. The Fourier transform in the time domain is defined by the eigenvalue decomposition of the Laplace matrix \cite{ft}, i.e., 
\begin{align}
    \bm{L}_d^k=\bm{U}\bm{\Lambda}\bm{U}^\top
\end{align}
where the eigenvalues $\lambda_j$ denote the frequency and the eigenvectors $\bm{u}_j$ denote the Fourier modes, respectively. Regarding $\bm{x}$ as a signal in $G_d^k$, we can define the Fourier transform as $\breve{\bm{x}}={\rm FT}(\bm{x})=\bm{U}^\top\bm{x}$ and the inverse Fourier transform as $\bm{x}={\rm IFT}(\breve{\bm{x}})=\bm{U}\breve{\bm{x}}$. Therefore, the convolution between $\bm{x}$ and the convolution kernel $g_\theta={\rm diag}(\theta)$ parameterized by $\theta\in \mathbb{R}^{|\bm{x}|}$ can be represented as
\begin{align}
    g_\theta\ast \bm{x}=\bm{U}g_\theta\bm{U}^\top\bm{x}
    \label{ftconv}
\end{align}
where $\odot$ is the Hadamard product.

The complexity of the convolution kernel in (\ref{ftconv}) can be extremely high since the eigenvalue decomposition requires a $O(n^3)$ computation overhead on global information. A typical approach to decrease the computation complexity is to approximate (\ref{ftconv}) by truncated Chebyshev polynomials of the first order according to \cite{gcn}, i.e., 
\begin{equation}
\begin{aligned}
    g_\theta\ast \bm{x}
    &\approx\bm{U}(\mathop{\sum}\limits_{c=0}^1\beta_cT_c(\tilde{\bm{\Lambda}}))\bm{U}^\top\bm{x}
    =\mathop{\sum}\limits_{c=0}^1\beta_cT_c(\bm{U}\tilde{\bm{\Lambda}}\bm{U}^\top)\bm{x}\\
    &=\beta_0T_0(\tilde{\bm{L}}_d^k)\bm{x}+\beta_1T_1(\tilde{\bm{L}}_d^k)\bm{x}
    =\beta_0\bm{x}+\beta_1(\bm{L}_d^k-\bm{I}_N)\bm{x}
\end{aligned}
\end{equation}
where $T_c$ denotes the $c$-th item in the Chebyshev polynomials with $T_0(x)=1$ and $T_1(x)=x$, $\beta_c$ is the $c$-th Chebyshev coefficient, $\bm{I}_N\in \mathbb{R}^{N\times N}$ is the identity matrix, and $\tilde{\bm{\Lambda}}_d^k=\frac{2}{\lambda_{max}}\bm{\Lambda}_d^k-\bm{I}_N$ is scaled by the largest eigenvalue of $\bm{L}_d^k$. Inspired by \cite{mgc}, we define the hyper-parameter $\epsilon=-\beta_1>0$ and let $\beta_0 = \beta_1+1$, then we have the bipartite graph convolution operation for the \textit{United Graph} $G_d^k$ as
\begin{align}
    g_\theta\ast \bm{x}=(\bm{I}_N-\epsilon\bm{L}_d^k)\bm{x}.
\end{align}

\subsubsection{Solution of convolution output}
Let the position matrix of $G_d^k$ be the feature matrix, and the output matrix after $L$-th iteration of bipartite graph convolution operation can be expressed as
\begin{align}
    \bm{\hat{X}}_d = g_\theta\ast g_\theta\ast \cdots \ast g_\theta\ast\bm{X}_d=(\bm{I}_N-\epsilon\bm{L}_d^k)^L\bm{X}_d
    \label{g}
\end{align}
This indicates that the bipartite graph convolution maintains the shape of the feature matrix as $\bm{X}_d,\bm{\hat{X}}_d\in\mathbb{R}^{N\times 2}$ are isomorphic. Note that $\bm{X}_d$ consists of the location vectors of UAVs, $\bm{\hat{X}}_d$ can also represent a position matrix. By extracting the first $N_R$ location vectors from the convolution output $\bm{\hat{X}}_d$ as the target locations for each remaining UAV, the final position matrix of the \textit{Remained Graph} $\hat{G}_R$ can be expressed as
\begin{equation}
    \bm{\hat{X}}_R=[\bm{I}_{N_R}\enspace\bm{0}_{N_D}]\bm{\hat{X}}_d=[\bm{\hat{p}}_{r_1},\bm{\hat{p}}_{r_2},...,\bm{\hat{p}}_{r_{N_R}}]^\top.
\end{equation}

The recovery process for the USNET involves each remaining UAV $u_{r_i}$ flying at full speed towards its target position $\bm{\hat{p}}_{r_i}$ and remaining stationary upon arrival. Consequently, the total recovery time depends on the maximum time required for the remaining drones to reach their respective target positions. Thus, problem (P1) can be reformulated as 
\begin{align}
    ({\rm P}2):\quad \mathop{\rm min}\limits_{\bm{\hat{X}}_R}\quad & T_{rc}=\mathop{\rm max}\limits_{r_i\in\mathcal{I}_R}\frac{\|\bm{\hat{p}}_{r_i}-\bm{p}_{r_i}(t_0)\|}{v_{max}}
    \label{p2}\\
    \quad {\rm s.t.}\quad & \hat{N}_S = 1
    \label{c2}
\end{align}
where $\hat{N}_S$ is the number of sub-nets in $\hat{G}_R$ based on the position matrix $\bm{\hat{X}}_R$.

\subsection{Framework of MDSG-GC}
Building on the solution to the CNS problem provided by the bipartite graph convolution, we present the overall architecture of MDSG-GC as shown in Fig., which comprises three components: the batch processing mechanism, the GCO backbone network, and the loss function design.

\begin{figure*}[!t]
\centerline{\includegraphics[width=\linewidth]{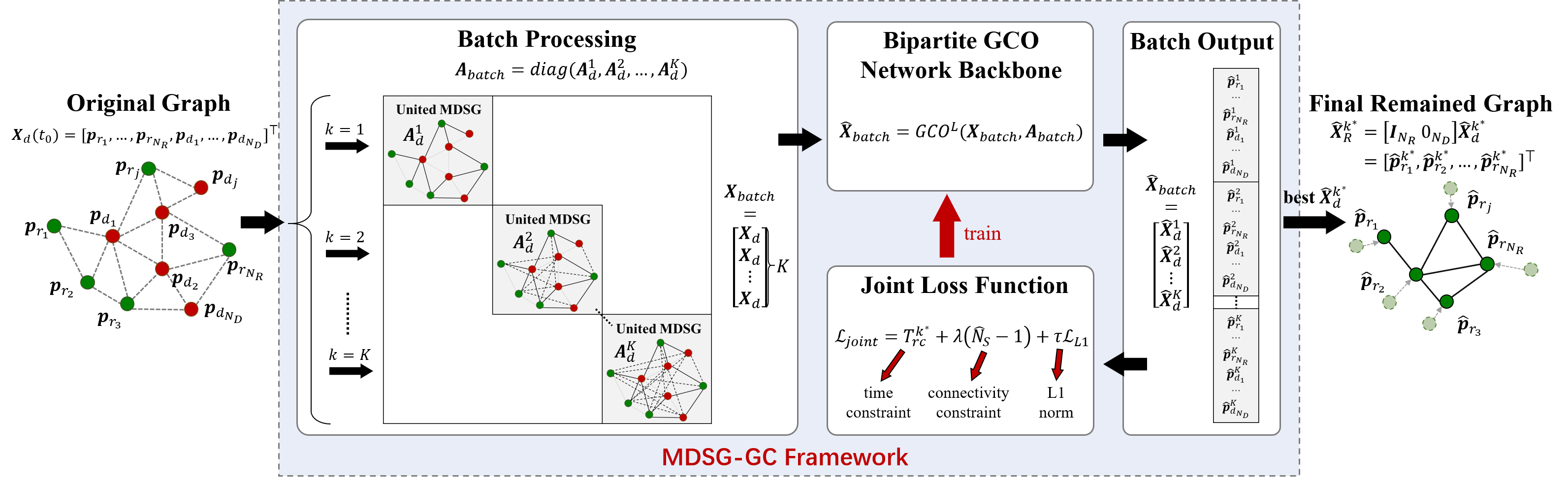}}
\caption{Overall framework of MDSG-GC. The framework includes three key components: a batch mechanism for input and output processing, a GCO backbone network for solution generation, and a joint loss function for training.}
\label{framework}
\end{figure*}

\subsubsection{Batch processing mechanism}

When constructing the MDSG $G_{d, r_i}^k$ and the \textit{United Graph} $G_d^k$, the hop number $k$ significantly influences the final topology. A larger $k$ provides nodes with a broader view of their neighboring topology, enabling the MDSG to capture more information from the original network. However, an excessively large $k$ can create a complex multi-hop differential relationship between too many damaged and remaining nodes, which can diminish the effectiveness of the damaged information in guiding localized recovery. Therefore, selecting an appropriate $k$ is crucial for the performance of the method. In our previous work \cite{demd}, the choice of $k$ was primarily based on network size and corruption scale, which generally yielded statistically satisfactory results. Nonetheless, this $k$ may not be optimal for all specific damage scenarios. Inspired by the batch processing used in graph convolutional networks \cite{gcn}, we propose a mechanism to address (P2) by evaluating different $k$ values in parallel.

The essence of graph convolution batching lies in combining multiple graphs into a new, disconnected graph by stitching them diagonally. For different hop numbers $k = 1,2,..., K$, we can generate a set of united MDSGs $\{G_d^1, G_d^2,..., G_d^K\}$ and their adjacency matrices $\{\bm{A}_d^1, \bm{A}_d^2,..., \bm{A}_d^K\}$. The input graph $G_{batch}$ is constructed by merging these united graphs with
\begin{align}
    \bm{X}_{batch}=[\underbrace{\bm{X}_d^\top,\bm{X}_d^\top,...,\bm{X}_d^\top}_K]^\top\\
    \bm{A}_{batch}={\rm diag}(\bm{A}_d^1, \bm{A}_d^2,..., \bm{A}_d^K)
\end{align}
where $\bm{X}_{batch}\in \mathbb{R}^{KN\times 2}$ denote the concatenated feature matrix, and the block-diagonal matrix $\bm{A}_{batch}\in\mathbb{R}^{KN\times KN}$ denote the adjacency matrix of $G_{batch}$. 

The batch processing mechanism eliminates the need to pre-define a single $k$ by directly providing solutions for different $k$ values. This allows us to determine the optimal $k*$ for each damage case. While available computational resources constrain the choice of batch size $K$, our subsequent analysis will demonstrate that the MDSG-GC model significantly reduces both pre-training overhead and storage requirements.

\subsubsection{Bipartite GCO network backbone}
To minimize $T_{rc}$ in (\ref{p2}), we extend the bipartite graph convolution operation $g_\theta$ into an $L$-layer graph convolutional network (GCN). In each graph convolutional layer, we introduce a weight matrix for feature linear transformation, which serves as the trainable parameter. Specifically, in the $l$-th layer, the feature matrix is processed by the GCO as
\begin{align}
    \bm{X}_{batch}^{l+1}&={\rm GCO}(\bm{X}_{batch}^l,\bm{A}_{batch})\\
    &=\sigma\left((\bm{I}_{KN}-\epsilon\bm{L}_{batch})\bm{X}_{batch}^l\bm{W}^l\right)
    \label{w}
\end{align}
where $\sigma(\cdot)$ denotes an activation function, ${\rm\bf X}^l\in \mathbb{R}^{KN\times d_s}$ is the feature matrix in the $l^{th}$ layer with a hidden dimension of $d_s$, and ${\rm\bf W}^l$ is a layer-specific trainable weight matrix. The Laplace matrix $\bm{L}_{batch}$ is computed based on $\bm{A}_{batch}$. In addition, the dropouts \cite{dropout} are applied between two layers to increase the generalization ability of the GCN.

The output feature matrix of the GCN is a batch output with
\begin{align}
    \hat{\bm{X}}_{batch}={\rm GCO}^L(\bm{X}_{batch},\bm{A}_{batch})=[\bm{\hat{X}}_d^{1\top},\bm{\hat{X}}_d^{2\top},...,\bm{\hat{X}}_d^{K\top}]^\top
\end{align}
where each $\bm{\hat{X}}_d^k=[\bm{\hat{p}}_{r_1}^k,...,\bm{\hat{p}}_{r_{N_R}}^k,\bm{\hat{p}}_{d_1}^k,...,\bm{\hat{p}}_{d_{N_D}}^k]^\top\in\mathbb{R}^{N\times 2}$ is a possible solution for (P2). The best solution is determine by calculate the recovery time $T_{rc}^k$ of different $k$, that is,
\begin{align}
    k^*=arg \mathop{\rm min}\limits_{k}T_{rc}^k=arg \mathop{\rm min}\limits_{k}\mathop{\rm max}\limits_{r_i\in\mathcal{I}_R}\frac{\|\bm{\hat{p}}_{r_i}^k-\bm{p}_{r_i}(t_0)\|}{v_{max}}
\end{align}
and the position matrix of the final \textit{Remained Graph} is given as
\begin{align}
    \bm{\hat{X}}_R^{k^*}=[\bm{I}_{N_R}\enspace\bm{0}_{N_D}]\bm{\hat{X}}_d^{k^*}=[\bm{\hat{p}}_{r_1}^{k^*},\bm{\hat{p}}_{r_2}^{k^*},...,\bm{\hat{p}}_{r_{N_R}}^{k^*}]^\top.
    \label{sl}
\end{align}
where $\bm{\hat{p}}_{r_i}^{k^*}$ denotes the target recovery location of the remaining node $u_{r_i}$.

Denote $T_{r_i,rc}=\frac{1}{v_{max}}\|\bm{\hat{p}}_{r_i}^{k^*}-\bm{p}_{r_i}(t_0)\|$ is the flying time of the remaining node $u_{r_i}$ from its initial location to the target recovery location. The velocities at each time step during the recovery process are calculated as
\begin{equation}
    \begin{aligned}
        \bm{\hat{v}}_{r_i}(t_l)=\begin{cases}
            v_{max}\frac{\bm{\hat{p}}_{r_i}^{k^*}-\bm{p}_{r_i}(t_0)}{\|\bm{\hat{p}}_{r_i}^{k^*}-\bm{p}_{r_i}(t_0)\|},\quad  &t_0\leq t_l < t_0+T_{r_i,rc}\\
            0,&otherwise
        \end{cases}
    \end{aligned}
\end{equation}
and the final solution of MDSG-GC for (P1) is 
\begin{align}
    \hat{\mathcal{V}}_R=\{\hat{\bm{V}}_R(t_l)|t_0\leq t_l<t_0+T_{rc}^{k^*}\}
\end{align}
where $\bm{\hat{V}}_R(t_l)=[\bm{\hat{v}}_{r_1}(t_l),\bm{\hat{v}}_{r}(t_l),...,\bm{\hat{v}}_{r_{N_R}}(t_l)]^\top$ is the velocity matrix at time $t_l$.

\subsubsection{Joint loss function}
Denote the loss function of the GCN as $\mathcal{L}(\bm{X}_{batch},\mathcal{W})$, where $\bm{X}_{batch}$ is the input feature matrix and $\mathcal{W}=\{\bm{W}^1,\bm{W}^2,...,\bm{W}^L\}$ is the set of layer weight matrices. Based on the solution in (\ref{sl}), we rewrite (P2) as
\begin{align}
    ({\rm P}2^\dagger):\quad \mathop{\rm min}\limits_{\hat{\bm{X}}_{batch},k^*}\quad & T_{rc}^{k^*}=\mathop{\rm min}\limits_{k}\mathop{\rm max}\limits_{r_i\in\mathcal{I}_R}\frac{\|\bm{\hat{p}}_{r_i}^k-\bm{p}_{r_i}(t_0)\|}{v_{max}}
    \label{p2s}\\
    \quad {\rm s.t.}\quad & \hat{N}_S^{k^*} - 1 \leq 0
\end{align}
where $T_{rc}$ is substituted by the recovery time of the best solution $T_{rc}^{k^*}$, and the constraint (\ref{c2}) is represented by the inequality $\hat{N}_S^{k^*} - 1 \leq 0$. Based on (P$2^\dagger$) we can form a basic loss function of the GCN as
\begin{align}
    \mathcal{L}(\bm{X}_{batch},\mathcal{W})=T_{rc}^{k^*}+\lambda (\hat{N}_S^{k^*} - 1)
    \label{loss1}
\end{align}
where the Lagrange multiplier $\lambda$ is set as a positive constant.

However, the current loss function in (\ref{loss1})  prioritizes only the best solution in each training epoch, leading to slow iterations and extensive training time for the remaining solutions with varying k values. To address this issue, it is crucial to introduce a constraint into the loss function to facilitate parallel solving. Given that the optimal solution for CNS recovery necessitates minimizing the distance flown by each UAV, incorporating an L1-norm can effectively capture this constraint. The L1-norm loss component can be expressed as 
\begin{align}
\mathcal{L}_{L1}=\frac{1}{K\cdot N_R}\mathop{\sum}\limits_{k=1}^K\mathop{\sum}\limits_{r_i\in \mathcal{I}_R}\|\bm{\hat{p}}_{r_i}^{k}-\bm{p}_{r_i}(t_0)\|.
\end{align}

The joint loss function for the GCN includes three components: the time constraint, the connectivity constraint, and the L1-norm. It is expressed as
\begin{align}
    \mathcal{L}_{joint}=T_{rc}^{k^*}+\lambda (\hat{N}_S^{k^*} - 1)+\tau\mathcal{L}_{L1}
\end{align}
where the multiplier $\tau$ is set as a positive coefficient to balance the gradients. To enhance the model's generalization capability, we incorporate the GradNorm \cite{gradnorm} mechanism into $\mathcal{L}_{joint}$. The values of $\lambda$ and $\tau$ are adaptively adjusted based on the gradients of the components, hence mitigating gradient vanishing issues and effectively boosting overall model performance.


\subsection{Theoretical Analysis of MDSG-GC}
To analyze MDSG-GC, we need to address whether the solution can guarantee connectivity recovery and assess its performance under worst-case scenarios. We first demonstrate that the bipartite graph convolution operation converges for the feature matrix $\bm{X}_d$ of the united MDSG $G_d^k$. Subsequently, we derive the theoretical upper bound for the recovery time, relying on the final convergent matrix of the bipartite graph convolution.

\subsubsection{Convergence of bipartite graph convolution}
For the convolution output $\bm{\hat{X}_d}$ in (\ref{g}), we can prove that the bipartite graph convolution is a contraction operation for each iteration.

\begin{proposition}
In the metric space of position matrices $\{\bm{X}_d\}\subseteq\mathbb{R}^{N\times 2}$, the bipartite GCO is a contraction operation when $0<\epsilon\leq\frac{1}{\|\bm{A}_d^k\|_\infty}$, and there exists and only exists one position matrix $\bm{\bar{X}}_d=[\bm{\bar{p}}_{d,1},\bm{\bar{p}}_{r_{d,2}},...,\bm{\bar{p}}_{d,N}]^\top$ such that
\begin{align}
    \mathop{\rm lim}\limits_{l\rightarrow\infty}(\bm{I}_N-\epsilon\bm{L}_d^k)^l\bm{X}_d=\bm{\bar{X}}_d.
\end{align}

Especially, $\bm{\bar{X}}_d\equiv[\bm{p}_c,\bm{p}_c,...,\bm{p}_c]^\top$ holds when the united MDSG $G_d^k$ is connected, where $\bm{p}_c=\frac{1}{N}\sum_{i\in\mathcal{I}}\bm{p}_{d,i}(t_0)$ denotes the central location of all UAVs in the original USNET at time $t_0$.
\label{prop-c2}
\end{proposition}

\begin{IEEEproof}
See Appendix B.
\end{IEEEproof}

Proposition \ref{prop-c2} demonstrates that, after a sufficient number of iterations, the batch processing output $\bm{\hat{X}}_d^k$ converges to $\bm{\bar{X}}_d^k$. Since the \textit{Remained Graph} constructed from $\bm{\bar{X}}_d\equiv[\bm{p}_c,\bm{p}_c,...,\bm{p}_c]^\top$ is fully connected, we need to find the batch size $K$ that ensure at least one connected $G_d^k$ with $k\leq K$, and determine a proper $\epsilon$ to meet the convergence constraint for each united MDSG with different $k$.

\subsubsection{Choice of $K$ and $\epsilon$}
As the number of nodes $N$ and the extent of damage $N_D$ increase, a larger $k$ is required to ensure that $G_d^k$ remains connected. However, the batch size $K$ limits the maximum value of $k$ with $k_{max}=K$ that the MDSG-GC can process. It can be easily observed that if $G_d^{k_i}$ is connected, then for every $k_i<k\leq K$, graph $G_d^k$ is connected.
Therefore, we only need to ensure that $G_d^K$ is connected in the worst-case scenario. Denote $H_{max}$ as the maximum hop number in the original USNET, we set the choice of $K$ as
\begin{equation}
    K = \lfloor \frac{H_{max}+1}{2} \rfloor.
    \label{K}
\end{equation}
where $\lfloor\cdot\rfloor$ denotes the floor function.

For the choice of $\epsilon$, note that $\|\bm{A}_d^k\|_\infty$ is positively correlated to the value of $k$, but with a limit of $\|\bm{A}_d^k\|_\infty<N$. To ensure that the convolution kernel $g_\theta$ remains isomorphic for different $k$ during the batch process, we set $\epsilon=\frac{1}{N}<\frac{1}{\|\bm{A}_d^k\|_\infty}$ for the bipartite graph convolution to meet the convergence constraint. With the $K$ in (\ref{K}), the MDSG-GC must provide a solution that guarantees connectivity restoration, hence the convergence holds.

\subsubsection{Upper bound of $T_{rc}^{k^*}$}
The final solution $\bm{\hat{X}}_d^{k^*}$ of MDSG-GC also approaches $\bm{\bar{X}}_d$ when $G_d^{k^*}$ is connected. 
The worst case of recovery refers to the solution as $\bm{\hat{X}}_R^{k^*}=[\bm{p}_c,...,\bm{p}_c]^\top$ that guarantees the restoration of connectivity and therefore gives an upper bound for the recovery time as
\begin{align}
    T_{rc}^{max}=sup\{T_{rc}^{k^*}\} = \frac{\mathop{\rm max}\limits_{r_j\in \mathcal{I}_R}\|\bm{p}_c-\bm{p}_{r_j}(t_0)\|}{v_{max}}.
\end{align}

In practical applications, the model often does not need to iterate to $\bm{\bar{X}}_d$ to find a suitable solution, making early termination essential. According to Appendix B, when the number of iterations $l$ is small, the location of each remaining UAV tends to aggregate around the center of its MDSG. This occurs because the bipartite graph convolution prevents nodes from being exposed to global information too early, allowing solutions to be derived from local multi-hop differential information. This characteristic contributes to the performance advantage of the MDSG-GC.

\section{Simulation Results}
To credibly demonstrate the advantages of the MDSG-based algorithm in CNS recovery, we conducted a series of comparative simulations and case study experiments\footnote{The source codes are available on \textit{https://github.com/lytxzt/Multi-hop-Differential-Topology-based-Algorithms-for-Resilient-Network-of-UAV-Swarm}}. The setup and results of these experiments are stated as follows.

\subsection{Simulation Setup}
\subsubsection{Inferences} 
The original USNET consisting of $N=200$ UAVs was randomly distributed in a $1000\times 1000{\rm m}^2$ two-dimension area. Based on the analysis in \cite{mgc}, we set the communication distance as $d_{tr}=120{\rm m}$. The maximum value of UAV velocity was defined as $v_{max}=10{\rm m/s}$ for ease of calculation, and each recovery step had a time interval $\Delta t=0.1{\rm s/step}$.
The damage scenarios were divided into cases with different numbers of destroyed UAVs where $N_D$ was set from 10 to 190, and each simulation of different $N_D$ was repeated 50 times randomly for the statistic results.
For the hyper-parameters in the framework of MDSG-GC, the number of layers in the GCN is $L=8$, and the hidden dimension is $d_s=512$. The learning rate is set to 0.0001 with an Adam optimizer, and the dropout ratio is set to 0.1 to alleviate overfitting.


\subsubsection{Metrics}
Three metrics for performance evaluation were presented in this paper. The \textit{recovery time} $T_{rc}$ is the most important metric for measuring CNS recovery methods. 
Moreover, we introduced the \textit{spatial-coverage ratio} and \textit{cumulative degree distribution} to measure the spatial dispersion and topology uniformity of the final recovered $\hat{G}_R$, respectively.
The \textit{spatial-coverage ratio} represents the ratio of the  area covered by $\hat{G}_R$ after restoration to the area covered by the original USNET, hence a larger ratio indicates a more dispersed spatial distribution that mitigates the over-aggregation. 
The \textit{cumulative degree distribution} $P_d$ denotes the proportion of remaining nodes with a degree number not greater than $d$ versus the total number of nodes in $\hat{G}_R$. A more uniform degree distribution leads to a network that is more robust against attacks \cite{distribution}.

\subsection{Determine $k$ of the MDSG-APF}
\begin{figure}[!t]
\centerline{\includegraphics[width=\linewidth, height=.73\linewidth]{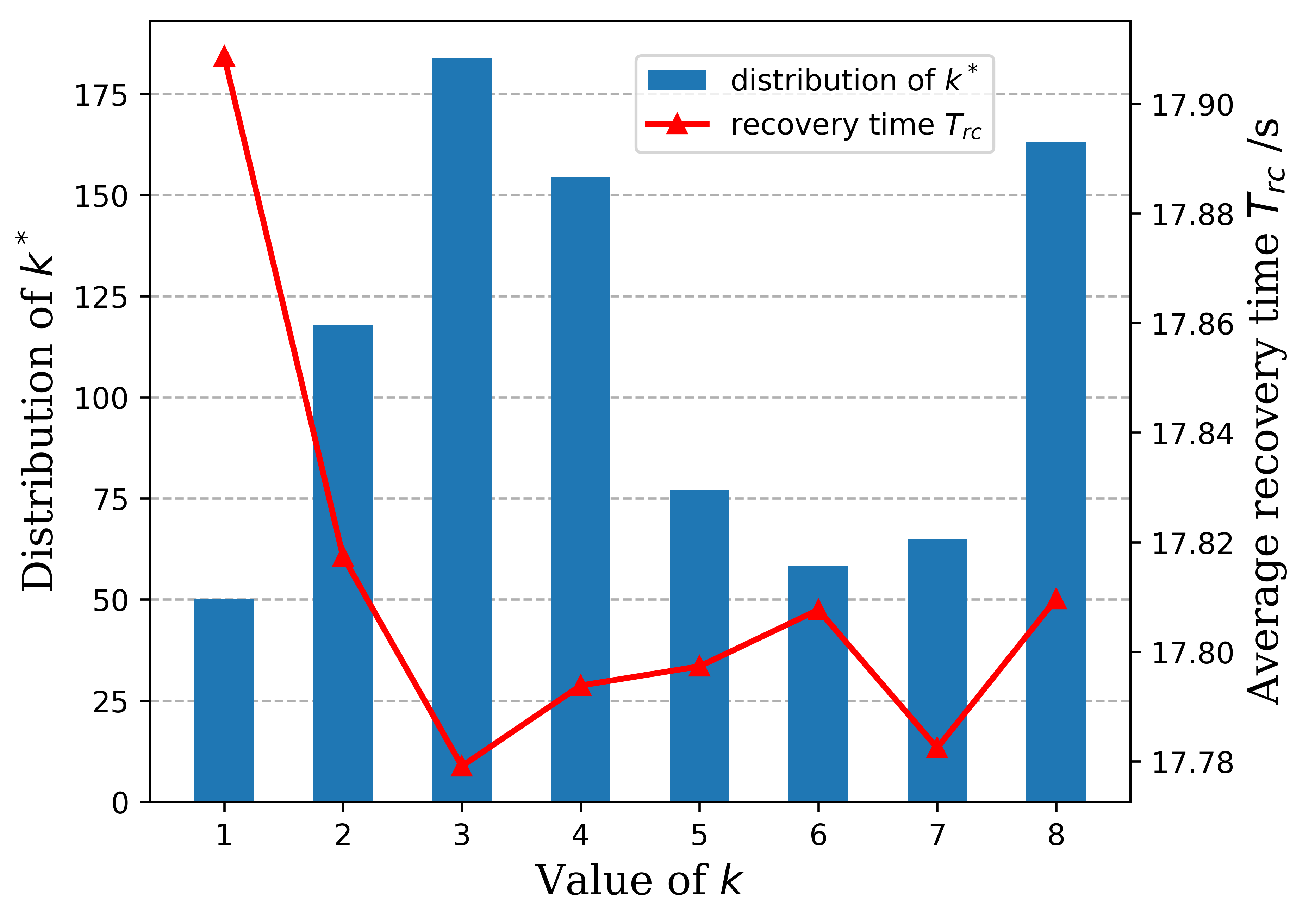}}
\caption{Average recovery time of MDSG-APF versus hop number $k$ and the distribution of best choice $k^*$.}
\label{khop}
\end{figure}

We randomly destructed 10 to 190 UAVs of the original USNET 50 times each to study the effectiveness of the hop number $k$ in the MDSG-APF. Fig.\ref{khop} plotted the average recovery time of the MDSG-APF with $k=1,2,3,4,5,6,7,8$ under different damage scales and count the frequency of different k values to the best result as the distribution of $k^*$. The results demonstrated that the average recovery time decreases first and then increases by $k$, meanwhile, the distribution of $k^*$ is concentrated in $k=3$, $k=4$, and $k=8$. Therefore, with the consideration of both the recovery efficiency and the computational overhead, we determine the value of $k=3$ for MDSG-APF.

\subsection{Batch processing of the MDSG-GC}
The batch processing mechanism offers two key advantages to MDSG-GC. First, it allows parallel computation of solutions $\bm{\hat{X}}_d^k$ with different $k$ values, optimizing the recovery path for each damage scenario. Second, it uses a unified input matrix framework, necessitating only one pre-trained model to adapt to various damage scenarios. This significantly reduces storage needs for pre-trained parameters and decreases the overhead of online iterative computing.

\begin{figure}[!t]
\centerline{\includegraphics[width=\linewidth]{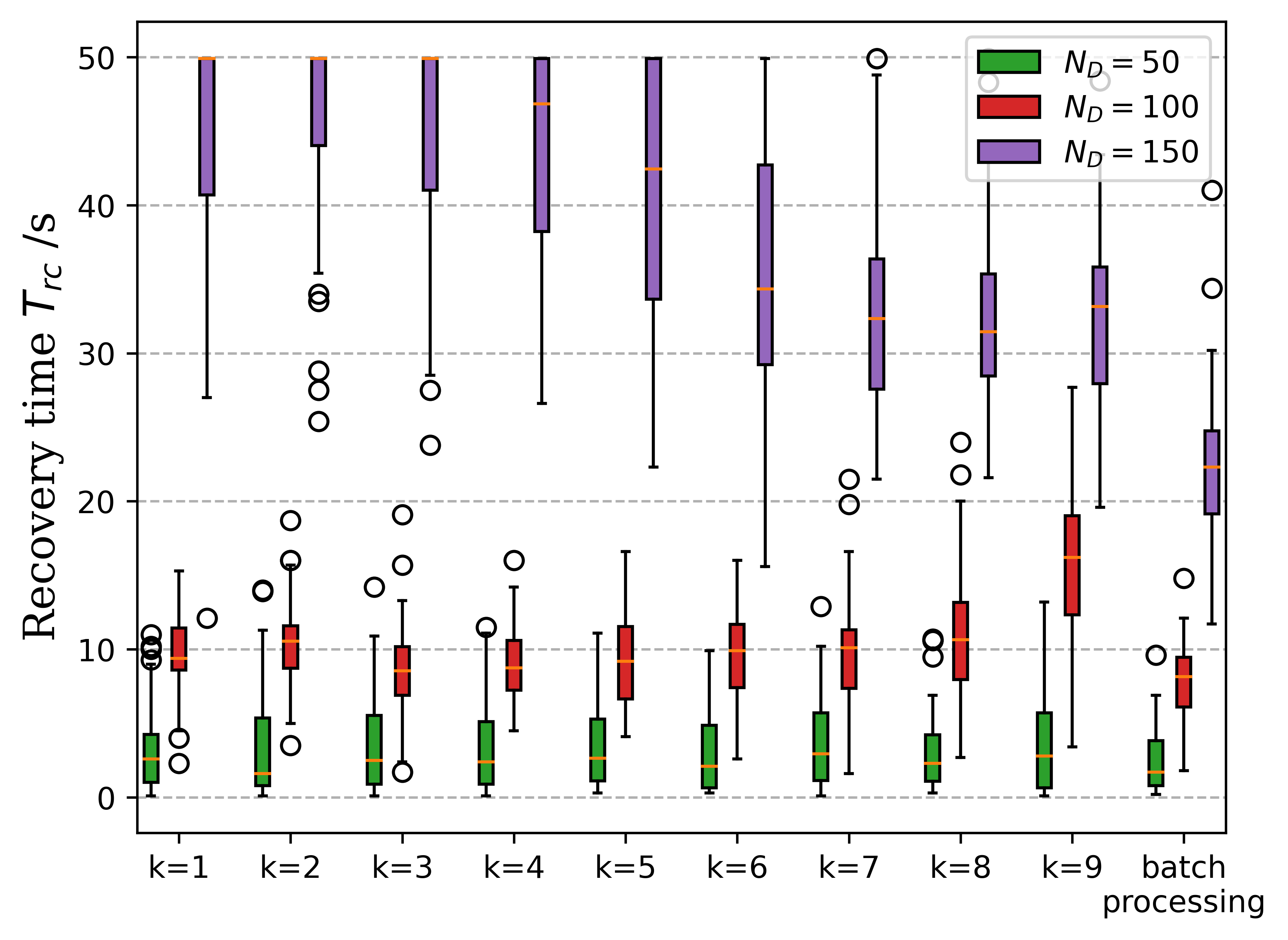}}
\caption{Recovery time of MDSG-GC under different $N_D$ with batch processing compared to specific $k$.}
\label{batch}
\end{figure}
\begin{figure}[!t]
\centerline{\includegraphics[width=\linewidth]{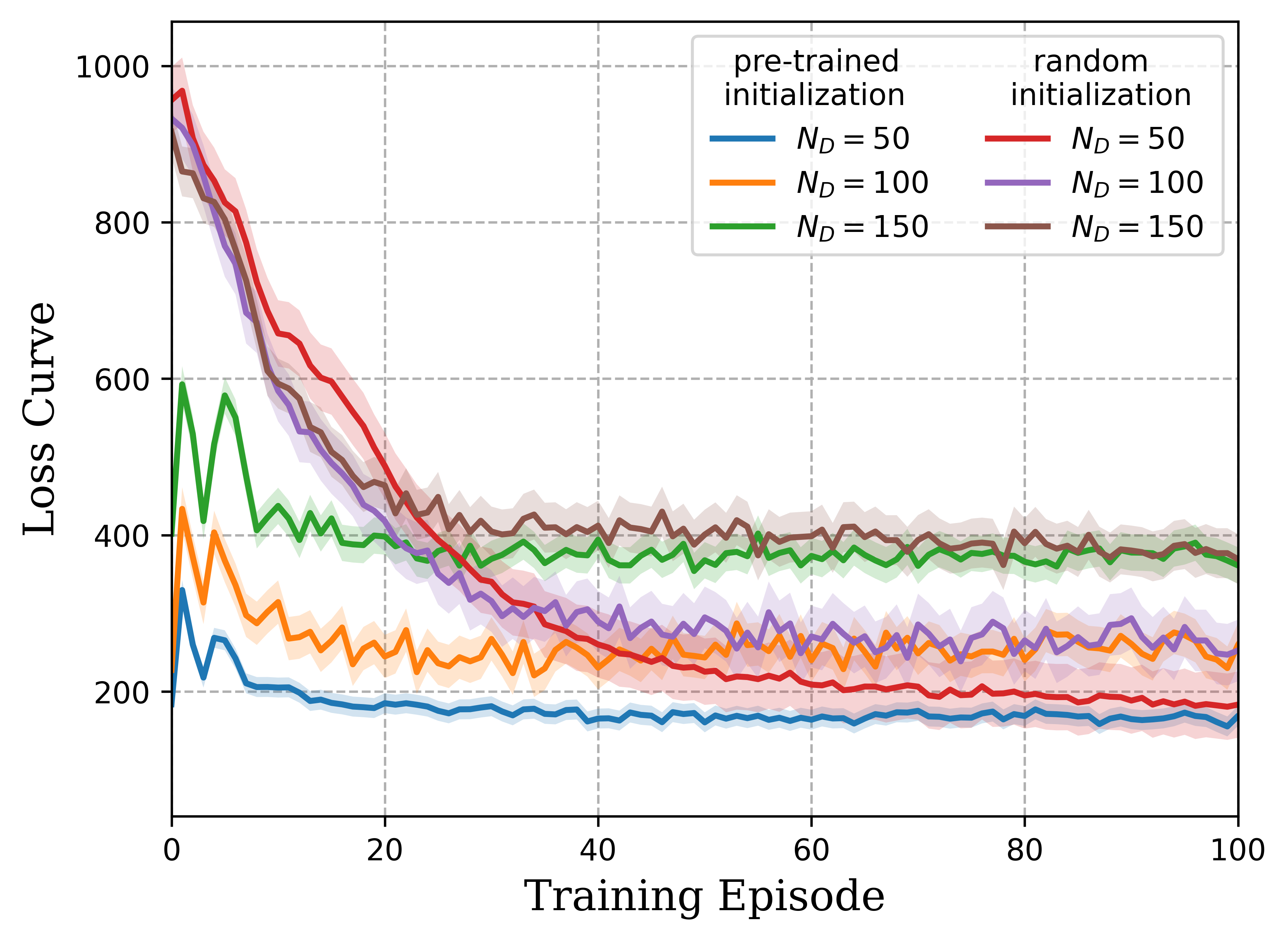}}
\caption{Loss curve of MDSG-GC training with $W$ initialized by pre-trained model and random values under different $N_D$.}
\label{loss}
\end{figure}

The recovery time of the MDSG-GC with batch processing and specific $k$ values are shown in Fig.\ref{batch}. The results demonstrate that the optimal value of specific $k$ corresponds to the scale of destructions, e.g., the best choice of $k$ is $k=3$ when $N_D=100$ while it becomes $k=9$ when $N_D=150$.
Compared with specific $k$ values, the MDSG-GC with batch processing had a shorter recovery time across all $N_D$. Although batch processing increases the number of parameters, the MDSG-GC has only a \textit{6.02Mb} model size that is manageable given the performance gains achieved.

The loss function curves of the MDSG-GC during the training process were plotted in Fig.\ref{loss}, where the weights $W$ in (\ref{w}) were initialized by pre-trained model and random parameters, respectively. The loss function curves of MDSG-GC initialized by the pre-trained model started from smaller loss values and achieved a smooth stage at the 20th episode. This means that the pre-trained model can calculate a solution for each CNS issue within a quite short time. Meanwhile, we can see the same pre-trained model can initialize the MDSG-GC under different $N_D$ with similar curve trends. Therefore, the MDSG-GC reduces the storage requirement to only one pre-trained model compared with CR-MGC, which requires 199 meta-learning models.

\begin{figure*}[!t]
\centering{
\subfloat[average recovery time $T_{rc}$]{\includegraphics[width=.475\linewidth]{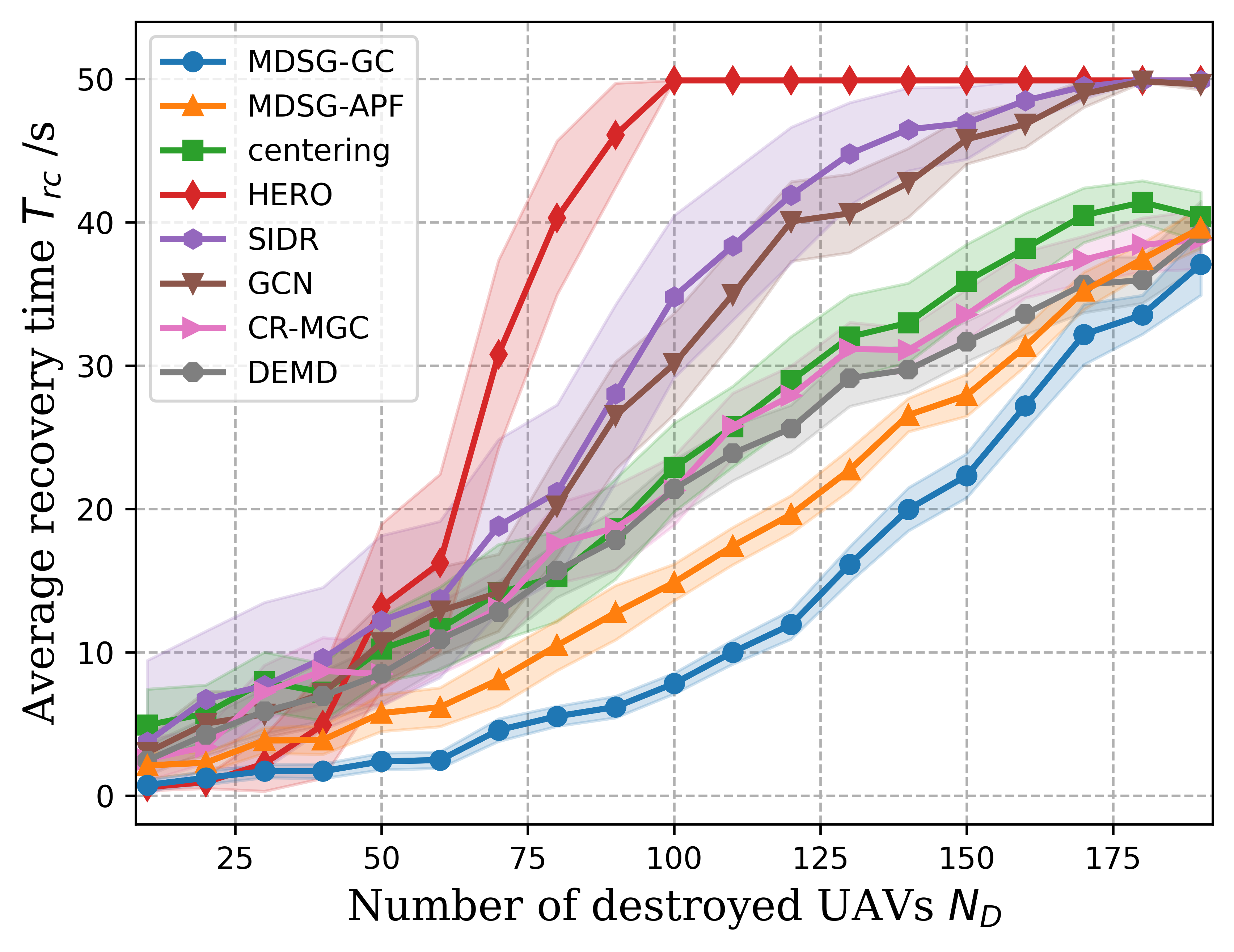}}\hspace{1em}
\subfloat[average spatial coverage ratios]{\includegraphics[width=.475\linewidth]{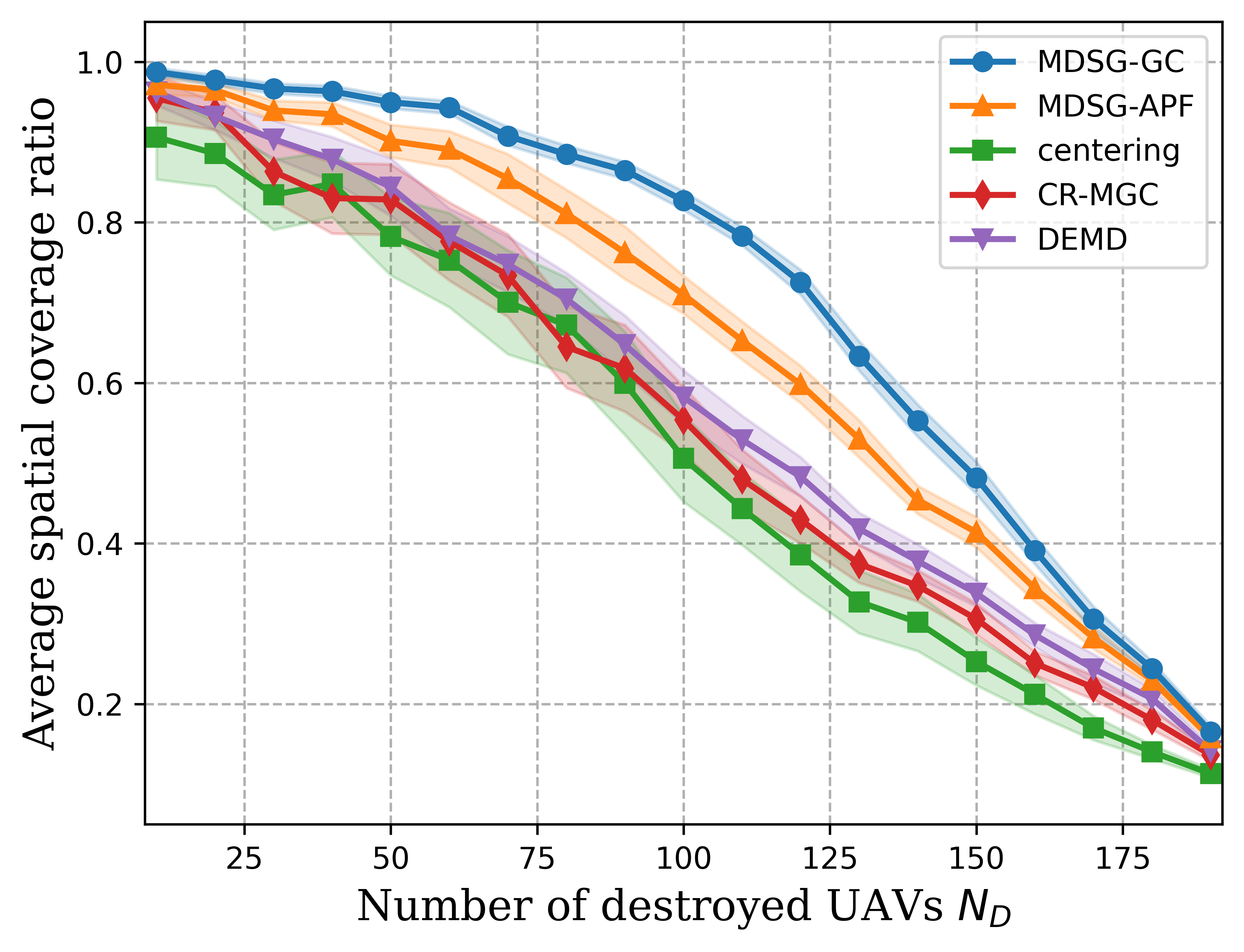}}\\
\subfloat[node degree distributions $P_d$ under $N_D=100$]{\includegraphics[width=.475\linewidth]{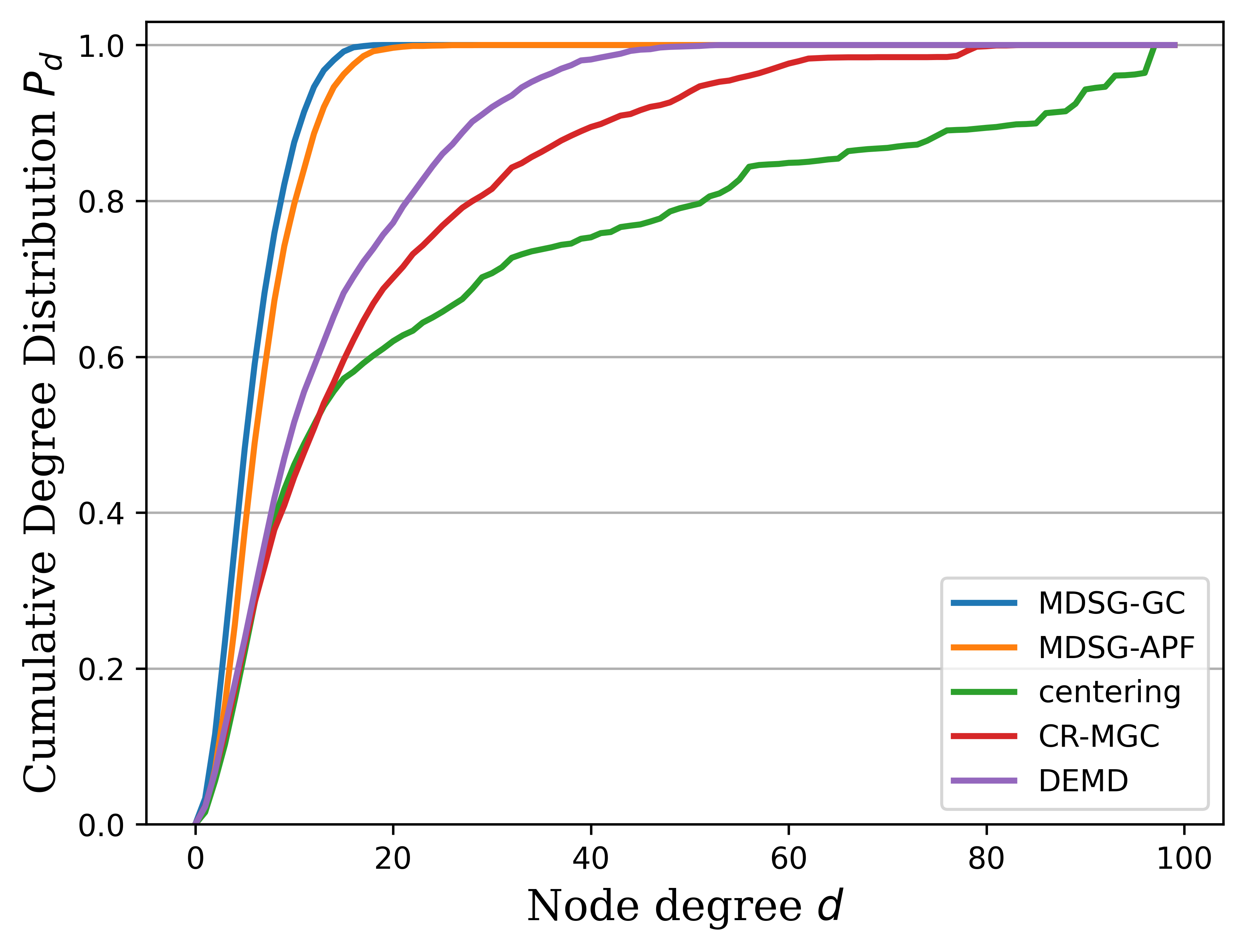}}\hspace{1em}
\subfloat[node degree distributions $P_d$ under $N_D=150$]{\includegraphics[width=.475\linewidth]{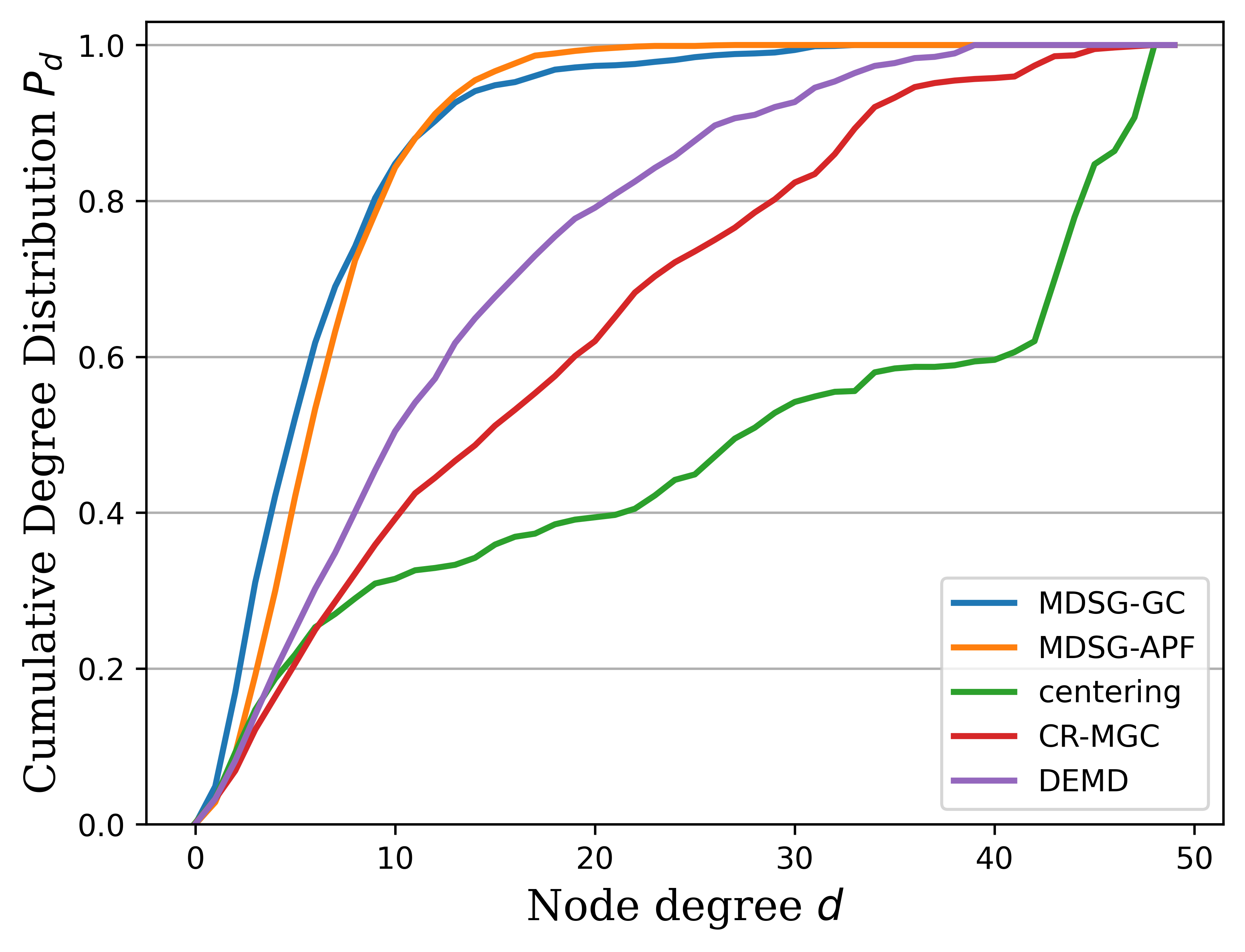}}}
\caption{Results evaluation of recovery time, spatial coverage ratios, and node degree distributions under different destructed UAV numbers with each approach.}
\label{res}
\end{figure*}

\subsection{Comparative Results Evaluation}
The average recovery time $T_{rc}$ of the MDSG-APF and MDSG-GC under damage of different $N_D$ were shown in Fig.\ref{res}a, where the values are statistically averaged from 50 times of random experiment. For comparisons, the performance of direct-centering, HERO \cite{hero}, SIDR \cite{sidr}, original GCN \cite{gcn}, CR-MGC \cite{mgc}, and our previous work DEMD \cite{demd} are also displayed. The shaded areas represent the confidential intervals of the average recovery time.
We can see that larger scales of damage $N_D$ require longer recovery time across all approaches. Once the damage exceeds 100 nodes, this growth trend accelerates significantly. 
Both the MDSG-APF and MDSG-GC show notable performance improvements; the MDSG-APF algorithm reduces an average of 23.2\% recovery time outperforming previous methods, while the MDSG-GC demonstrates even greater advantages with an average 50.2\% reduction on recovery time. For the half-damaged scenario with $N_D=100$, the average $T_{rc}$ of MDSG-GC is cut by 63.4\%, i.e., 10.7 seconds compared to CR-MGC and DEMD.

To evaluate the effectiveness of mitigating over-aggregation, we illustrated the spatial coverage ratio curve under varying $N_D$ of each approach in Fig.\ref{res}b. The results indicate that the coverage ratio decreases as $N_D$ increases since fewer remaining UAVs tend to cluster more densely. 
Notably, the two MDSG-based algorithms consistently achieved higher coverage ratios than other approaches, showing an average improvement of 14.7\% by MDSG-APF and 27.1\% by MDSG-GC outperforming other approaches. Under the half-damaged scenario $N_D=100$, the covered area of the recovered USNET by MDSG-GC is increased by $2.2\times 10^5 m^2$ compared with DEMD.
This demonstrates that the MDSG information can reconstruct the USNET with better spatial dispersion, effectively alleviating the issue of over-aggregation.

The cumulative node-degree distributions under 100 and 150 destroyed UAVs were illustrated in Fig.\ref{res}c and Fig.\ref{res}d. It is manifest that the degree distributions of the MDSG-based methods were mainly concentrated on smaller degrees. Moreover, the MDSG-APF and MDSG-GC represented the most concentrated distributions in the low-degree region with a faster-increased curve. Hence, the final network recovered via the MDSG has better topology uniformity, and the absence of large-degree nodes also improves the swarm resilience.

\subsection{Case Study of CNS Issue}

\begin{figure*}[!t]
\vspace{-1em}
\centering{
\subfloat[the number of sub-nets versus time]{\includegraphics[width=.31\linewidth]{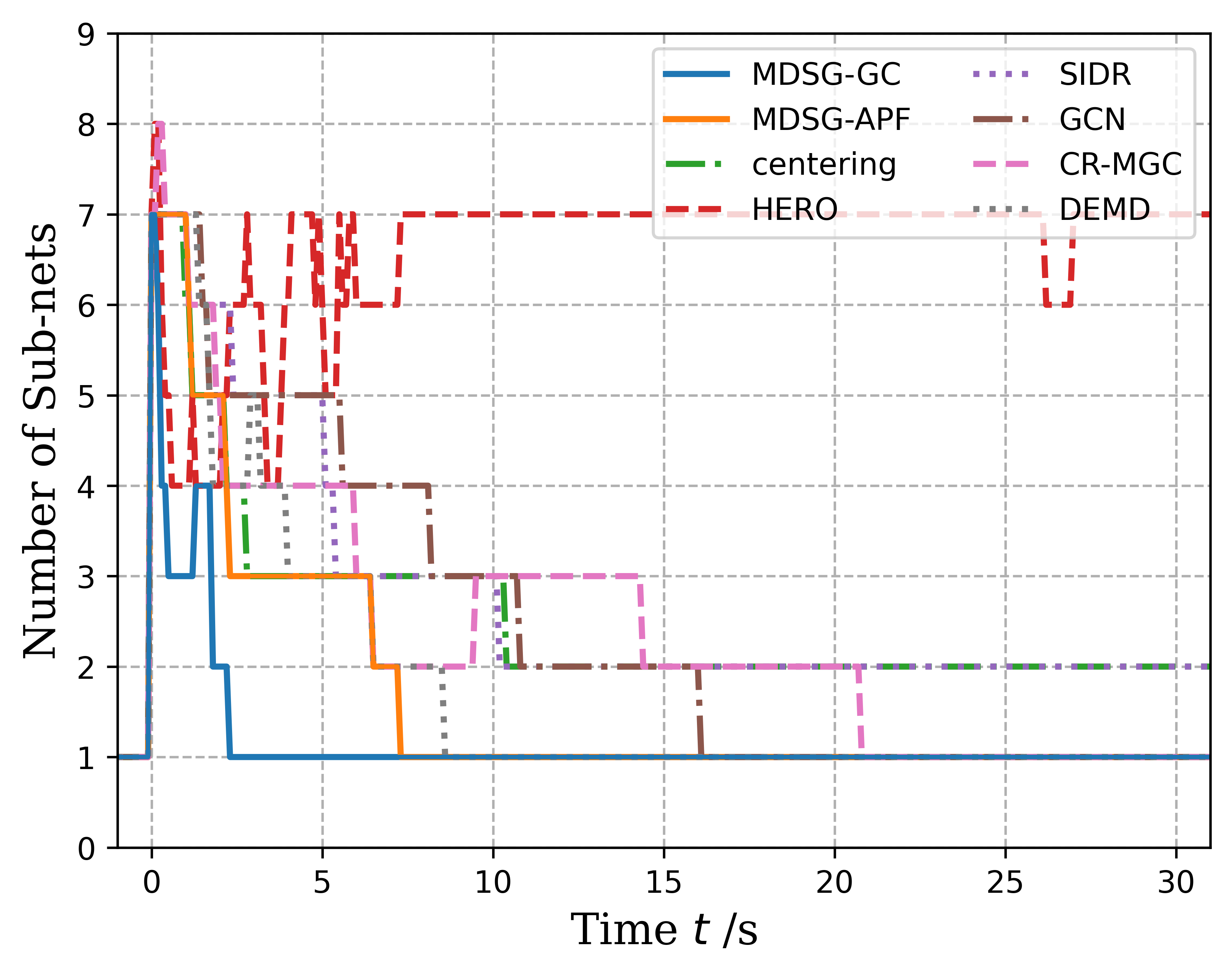}}\hspace{0.3em}
\subfloat[recovery trajectories by MDSG-APF]{\includegraphics[width=.34\linewidth]{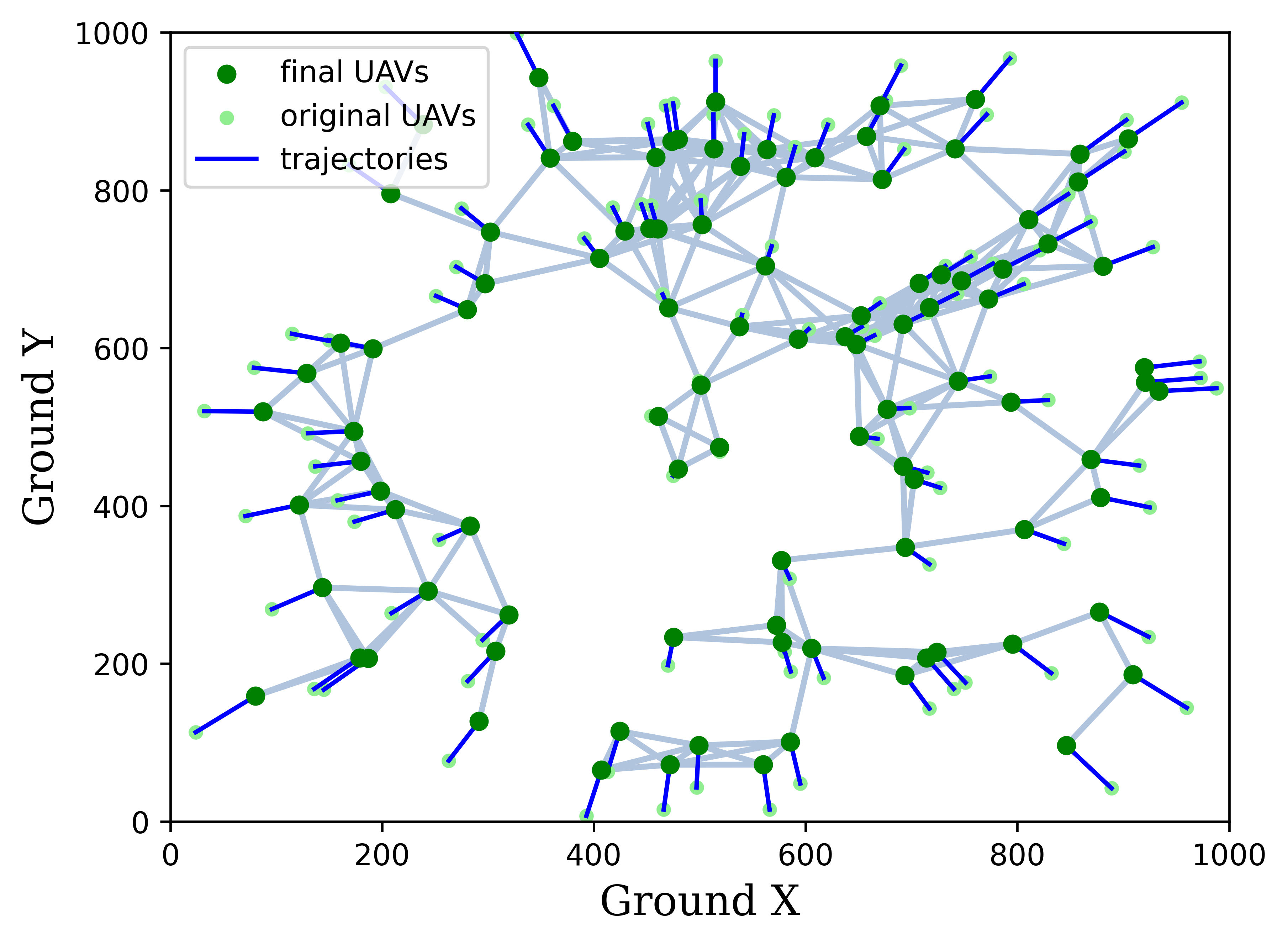}}\hspace{-0.5em}
\subfloat[recovery trajectories by MDSG-GC]{\includegraphics[width=.34\linewidth]{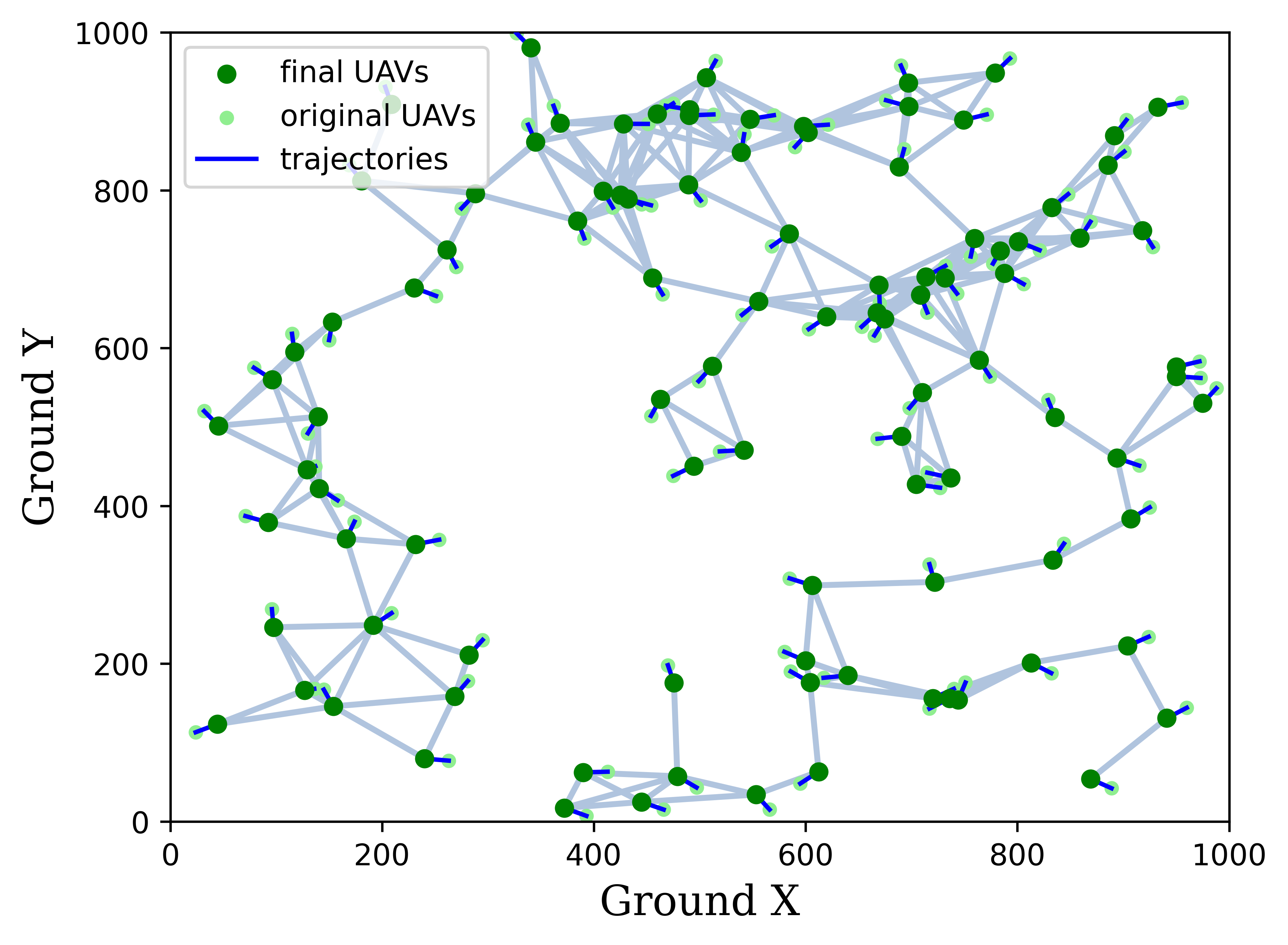}}}
\captionsetup{justification=raggedright, singlelinecheck=false}
\caption{Case study of CNS issue with 100 UAVs get destructed and recovered by different approaches.}
\label{case}
\end{figure*}

We randomly destruct $N_D=100$ UAVs of the original USNET to show the recovery process under the MDSG-based algorithms intuitively. In this case, the remaining USNET was divided into 7 disconnected sub-nets with different sizes, as shown in Fig.\ref{cha}a. The numbers of sub-nets during the process recovered by different algorithms were illustrated in Fig.\ref{case}b. It takes only 7.4s for MDSG-APF and 2.4s for MDSG-GC to reconstitute the connectivity of the \textit{Remained Graph}, while other algorithms require much more recovery time.
The recovery trajectories by MDSG-APF and MDSG-GC were plotted in Fig.\ref{case}c and Fig.\ref{case}d, respectively. The recovered USNETs showed wider spatial distributions and more uniform topologies, hence improving communication coverage and alleviating the problem of local over-aggregation.

\subsection{Time Consuming Comparisons}
\begin{figure}[!t]
\centerline{\includegraphics[width=.95\linewidth]{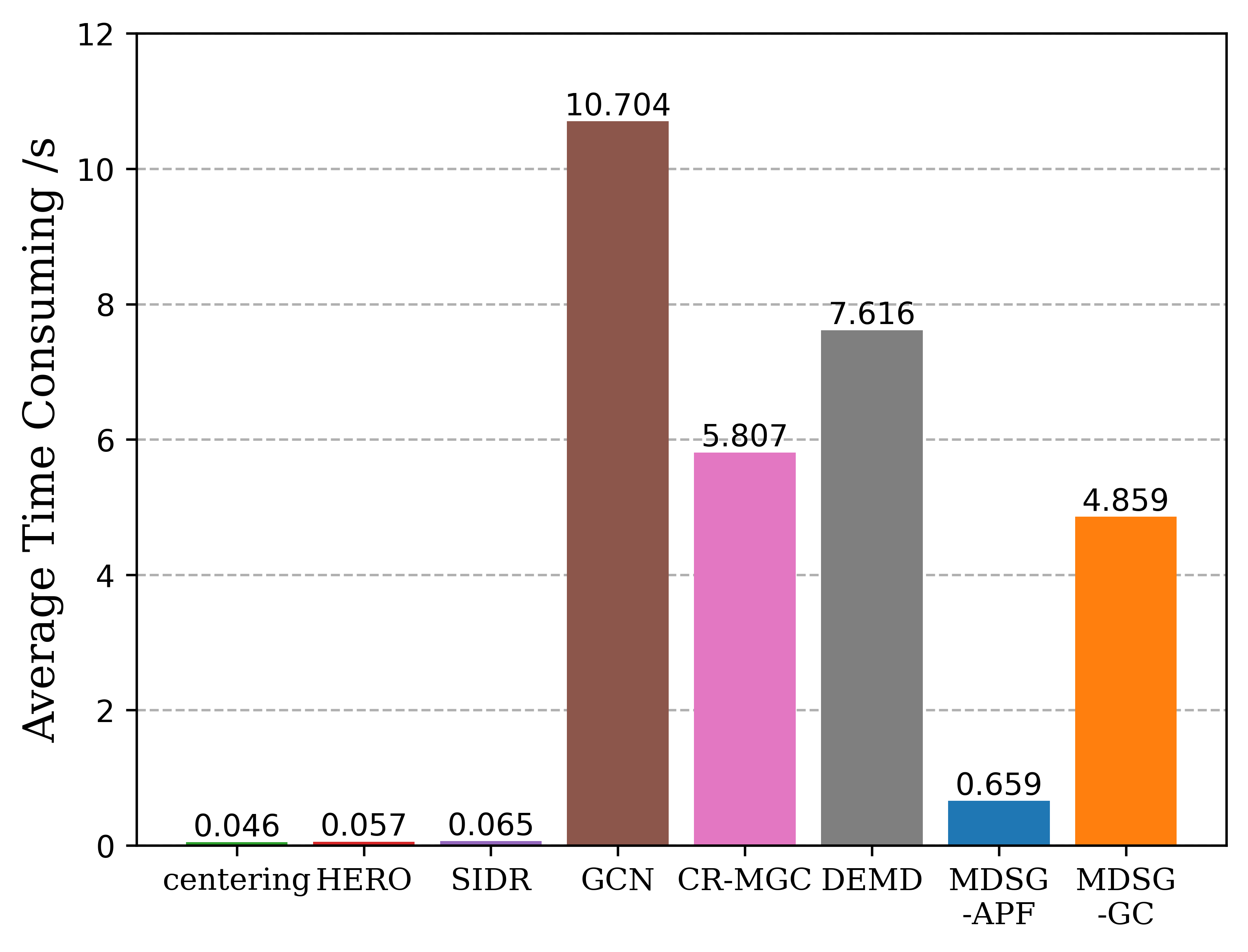}}
\caption{Average time consumption of different algorithms.}
\label{time}
\end{figure}

The average time consumption of different algorithms is compared in Fig.\ref{time}. It is worth noting that the centering, HERO, and SIDR require online calculation on every execution step, hence the entire time consumption of these algorithms is related to their recovery process. The other algorithms only need to calculate the recovery solution when the damage occurs. We can see that our MDSG-APF achieves a significant margin at time consumption, while the MDSG-GC costs less iteration time compared with other GCN-based algorithms. This indicates that the MDSG-based algorithms have acceptable time costs for addressing CNS issues.

\section{Conclusion}
In this paper, we studied the CNS issue of the USNET under massive damage. Specifically, we constructed the MDSGs for the remaining nodes to represent the destruction information of multi-hop neighboring nodes. We also developed two algorithms to plan the recovery trajectories based on MDSG for USNETs with low and high intelligence respectively. For the  USNETs with low intelligence, we designed the MDSG-APF algorithm to calculate the velocities for recovery. Meanwhile, for the USNETs with advanced intelligence, we proposed the MDSG-GC algorithm based on a batch-processing framework with a novel bipartite graph convolution operation to generate the final recovery positions for the remaining nodes. Statistical results demonstrated that the proposed MDSG-APF and MDSG-GC algorithms can reconstitute the connectivity within a shorter time with significant improvements in both spatial coverage ratio and topology degree distribution. The simulation results also showed that the batch-processing mechanism can enhance the performance of MDSG-GC while reducing the training time with the pre-trained model.


{\appendices
\section{Proof of Proposition \ref{prop-t}}
Since the velocity solution given by the MDSG-APF algorithm is constant during the recovery process, the recovery time $T_{rc}$ is positively correlated with the moving distance of each remaining node. Therefore, the upper bound of $T_{rc}$ refers to the worst-case scenario where the remaining nodes need to fly the farthest distance for recovery. According to Fig.\ref{velocity}, the final moving distance $d_{r_i}$ of each $u_{r_i}$ is calculated as
\begin{equation}
\begin{aligned}
    d_{r_i} &= \|\hat{\bm{p}}_{r_i}-\bm{p}_{r_i}(t_0)\|\\
    &= \|(\bm{p}_a-\bm{p}_{r_i}(t_0))+\alpha_{r_i}(\bm{p}_{d,r_i}-\bm{p}_{r_i}(t_0))\|\\
    &= \|\bm{v}_{a,r_i}(t_0) + \alpha_{r_i}\cdot\bm{v}_{d,r_i}(t_0)\|=\|\bm{v}_{r_i}(t_0)\|
\end{aligned}
\end{equation}
where $\bm{v}_{a,r_i}(t_0)$ and $\bm{v}_{d,r_i}(t_0)$ are given by (\ref{va}) and (\ref{vd}), respectively.

The time required to move the distance of $d_{r_i}$ for each $u_{r_i}$ is then represented as
\begin{equation}
\begin{aligned}
    T_{rc,r_i} =& \frac{d_{r_i}}{\|\hat{\bm{v}}_{r_i}(t_0)\|}
    = \frac{\|\bm{v}_{r_i}(t_0)\|}{\|\frac{v_{max}}{\mathop{\rm max}\limits_{r_j\in \mathcal{I}_R}\|\bm{v}_{r_j}(t_0)\|}\cdot\bm{v}_{r_i}(t_0)\|} \\
    =& \frac{\mathop{\rm max}\limits_{r_j\in \mathcal{I}_R}\|\bm{v}_{r_j}(t_0)\|}{v_{max}}\\
    =&\frac{\mathop{\rm max}\limits_{r_j\in \mathcal{I}_R}\|(\bm{p}_a-\bm{p}_{r_j}(t_0))+\alpha_{r_j}(\bm{p}_{d,r_j}-\bm{p}_{r_j}(t_0))\|}{v_{max}}\\
    \leq& \frac{\mathop{\rm max}\limits_{r_j\in \mathcal{I}_R}(\|\bm{p}_a-\bm{p}_{r_j}(t_0)\|+\alpha_{r_j}\|\bm{p}_{d,r_j}-\bm{p}_{r_j}(t_0)\|)}{v_{max}}\\
    =& \frac{\mathop{\rm max}\limits_{r_j\in \mathcal{I}_R}\|\bm{p}_a-\bm{p}_{r_j}(t_0)\|+\frac{d_{tr}}{2}}{v_{max}} = T_{rc,r_i}^{max}.
\end{aligned}
\end{equation}
where $T_{rc,r_i}^{max}=sup\{T_{rc,r_i}\}$ is the upper bound of recovery time for $u_{r_i}$, and $T_{rc,r_i}=T_{rc,r_i}^{max}$ if and only if $\bm{p}_a-\bm{p}_{r_j}(t_0)$ has the same direction as $\bm{p}_{d,r_j}-\bm{p}_{r_j}(t_0)$.

The recovery time of the entire network is required to consider the recovery time of every remaining node $u_{r_i}$, hence we have $T_{rc}=\mathop{\rm max}\limits_{r_i\in \mathcal{I}_R}T_{rc,r_i}$.
Note that $T_{rc,r_i}^{max}$ is independent of $u_{r_i}$, we can determine the upper bound for the entire recovery time as
\begin{equation}
\begin{aligned}
    T_{rc}^{max}=&sup\{T_{rc}\}
    =sup\{\mathop{\rm max}\limits_{r_i\in \mathcal{I}_R}T_{rc,r_i}\}\\
    =&\mathop{\rm max}\limits_{r_i\in \mathcal{I}_R}sup\{T_{rc,r_i}\}
    =\mathop{\rm max}\limits_{r_i\in \mathcal{I}_R}T_{rc,r_i}^{max}\\
    =&T_{rc,r_i}^{max}=\frac{\mathop{\rm max}\limits_{r_j\in \mathcal{I}_R}\|\bm{p}_a-\bm{p}_{r_j}(t_0)\|+\frac{d_{tr}}{2}}{v_{max}}.
\end{aligned}
\end{equation}

Assuming that the node $u_{r_m}={\rm arc} \mathop{\rm max}\limits_{r_j\in \mathcal{I}_R}\|\bm{p}_a-\bm{p}_{r_j}(t_0)\|$ is the farthest remaining node away from $\bm{p}_a$, the worst-case damage scenario is then described as $\bm{p}_a-\bm{p}_{r_m}(t_0)$ of node $u_{r_m}$ happens to have the same direction as $\bm{p}_{d,r_m}-\bm{p}_{r_m}(t_0)$.
This completes the proof.

\section{Proof of Proposition \ref{prop-c2}}
In the metric space of position matrices $\{\bm{X}_d\}\subseteq\mathbb{R}^{N\times 2}$, we can represent the matrix form of bipartite GCO $\bm{X}_d^{l+1}=g_\theta\ast \bm{X}_d^l=(\bm{I}_N-\epsilon\bm{L}_d^k)\bm{X}_d^l$ as
\begin{align}
    \begin{pmatrix}
    x_{d,1}^{l+1} & y_{d,1}^{l+1} \\
    x_{d,2}^{l+1} & y_{d,2}^{l+1} \\
    \vdots & \vdots \\
    x_{d,N}^{l+1} & y_{d,N}^{l+1}
    \end{pmatrix}=\begin{pmatrix}
    l_{d,11}^k &\!\!\!\! l_{d,12}^k &\!\!\!\! \ldots &\!\!\!\! l_{d,1N}^k \\
    l_{d,21}^k &\!\!\!\! l_{d,22}^k &\!\!\!\! \ldots &\!\!\!\! l_{d,2N}^k \\
    \vdots &\!\!\!\! \vdots &\!\!\!\! \ddots &\!\!\!\! \vdots \\
    l_{d,N1}^k &\!\!\!\! l_{d,N2}^k &\!\!\!\! \ldots &\!\!\!\! l_{d,NN}^k
    \end{pmatrix}\begin{pmatrix}
    x_{d,1}^l & y_{d,1}^l \\
    x_{d,2}^l & y_{d,2}^l \\
    \vdots & \vdots \\
    x_{d,N}^l & y_{d,N}^l
    \end{pmatrix}
\end{align}
where $x_{d,i}$ and $y_{d,i}$ denote the coordinate component of $X$ and $Y$ axis from the position vector $\bm{p}_{d,i}$ of UAV $u_{d,i}$, and $l_{d,ij}^k$ is the element of $\bm{I}_N-\epsilon\bm{L}_d^k$. Denote $d_{d,i}^k=\sum_{j=1}^Na_{d,ij}^k$ as the degree of node $u_{d,i}$ from the united MDSG $G_d^k$, we have
\begin{align}
\mathop{\sum}\limits_{i=1}^Nl_{d,ij}^k=1+\epsilon d_{d,i}^k-\epsilon\mathop{\sum}\limits_{j=1}^Na_{d,ij}^k=1.
\end{align}

The bipartite GCO can guarantee the invariance of column sums, i.e.,
\begin{align}
\mathop{\sum}\limits_{i=1}^Nx_{d,i}^{l+1}=\mathop{\sum}\limits_{i=1}^N\mathop{\sum}\limits_{j=1}^Nl_{d,ij}^kx_{d,j}^l=\mathop{\sum}\limits_{j=1}^N(x_{d,j}^l\mathop{\sum}\limits_{i=1}^Nl_{d,ij}^k)=\mathop{\sum}\limits_{j=1}^Nx_{d,j}^{l}
\end{align}
and the equality $\sum_{i=1}^Ny_{d,i}^{l+1}=\sum_{j=1}^Ny_{d,j}^l$ can be proved in the same manner.

Therefore, the bipartite GCO is closed for the metric space $\{\bm{X}_d|\sum_{i=1}^Nx_{d,i}=X_{sum},\sum_{i=1}^Ny_{d,i}=Y_{sum}\}$, which means every position matrix $\bm{X}_d^l$ share a same central location $\bm{p}_c=\frac{1}{N}\sum_{i=1}^N\bm{p}_{d,i}(t_0)=[\frac{1}{N}X_{sum},\frac{1}{N}Y_{sum}]^\top$ from $\bm{X}_d(t_0)$.

Then we use the \textit{Contraction Mapping Principle} to prove the bipartite GCO is a contraction operation and the position matrix $\bm{X}_d$ is convergent to a certain matrix $\bm{\bar{X}}_d$. We first define the distance between two position matrices $\bm{X}_d^a$ and $\bm{X}_d^b$ as 
\begin{equation}
\begin{aligned}
    d(\bm{X}_d^a,\bm{X}_d^b)&=\|\bm{X}_d^a-\bm{X}_d^b\|_\infty\\
    &=\mathop{\rm max}\limits_{1\leq i\leq N}\{|x_{d,i}^a-x_{d,i}^b|+|y_{d,i}^a-y_{d,i}^b|\}
\end{aligned}
\end{equation}

The distance between the bipartite GCO of $\bm{X}_d^a$ and $\bm{X}_d^b$ is then calculated as
\begin{equation}
\begin{aligned}
    d(g_\theta\ast \bm{X}_d^a,g_\theta\ast \bm{X}_d^b)&=\|g_\theta\ast \bm{X}_d^a-g_\theta\ast \bm{X}_d^b\|_\infty\\
    &=\|(\bm{I}_N-\epsilon\bm{L}_d^k)(\bm{X}_d^a-\bm{X}_d^b)\|_\infty
\end{aligned}
\end{equation}

Since the infinity norm of matrix $\|\cdot\|\infty$ has the sub-multiplicative property, we can get
\begin{equation}
\begin{aligned}
    d(g_\theta\ast \bm{X}_d^a,g_\theta\ast \bm{X}_d^b)\leq \|\bm{I}_N-\epsilon\bm{L}_d^k\|_\infty\|\bm{X}_d^a-\bm{X}_d^b\|_\infty\\
    =\mathop{\rm max}\limits_{1\leq i\leq N}\{|1-\epsilon d_{d,i}^k|+\mathop{\sum}\limits_{j=1}^N\epsilon a_{d,ij}^k\}\|\bm{X}_d^a-\bm{X}_d^b\|_\infty
\end{aligned}
\end{equation}

Notice that $d_{d,ij}^k\leq\|\bm{A}_d^k\|_\infty$, when $0<\epsilon\leq\frac{1}{\|\bm{A}_d^k\|_\infty}$, there is
\begin{align}
    1-\epsilon d_{d,i}^k\geq1-\epsilon\|\bm{A}_d^k\|_\infty\geq 1-\frac{1}{\|\bm{A}_d^k\|_\infty}\|\bm{A}_d^k\|_\infty=0
\end{align}
hence we have
\begin{equation}
\begin{aligned}
    d(g_\theta&\ast \bm{X}_d^a,g_\theta\ast \bm{X}_d^b)\\
    &\leq\mathop{\rm max}\limits_{1\leq i\leq N}\{|1-\epsilon d_{d,i}^k|+\mathop{\sum}\limits_{j=1}^N\epsilon a_{d,ij}^k\}\|\bm{X}_d^a-\bm{X}_d^b\|_\infty\\
    &=\mathop{\rm max}\limits_{1\leq i\leq N}\{1-\epsilon d_{d,i}^k+\mathop{\sum}\limits_{j=1}^N\epsilon a_{d,ij}^k\}\|\bm{X}_d^a-\bm{X}_d^b\|_\infty\\
    &=\mathop{\rm max}\limits_{1\leq i\leq N}\{1-\epsilon \mathop{\sum}\limits_{j=1}^Na_{d,ij}^k+\mathop{\sum}\limits_{j=1}^N\epsilon a_{d,ij}^k\}\|\bm{X}_d^a-\bm{X}_d^b\|_\infty\\
    &=\mathop{\rm max}\limits_{1\leq i\leq N}\{1\}\|\bm{X}_d^a-\bm{X}_d^b\|_\infty\\
    &=d(\bm{X}_d^a,\bm{X}_d^b)
    \label{dis}
\end{aligned}
\end{equation}

The condition for (\ref{dis}) to take the equal sign is
\begin{align}
    \|(\bm{I}_N-\epsilon\bm{L}_d^k)(\bm{X}_d^a-\bm{X}_d^b)\|_\infty=\|\bm{I}_N-\epsilon\bm{L}_d^k\|_\infty\|\bm{X}_d^a-\bm{X}_d^b\|_\infty
\end{align}
Given the proof of the sub-multiplicative property of $\|\cdot\|\infty$ as
\begin{equation}
\begin{aligned}
    &\|(\bm{I}_N-\epsilon\bm{L}_d^k)(\bm{X}_d^a-\bm{X}_d^b)\|_\infty\\
    &=\mathop{\rm max}\limits_{1\leq i\leq N}\{|\mathop{\sum}\limits_{j=1}^Nl_{d,ij}^k(x_{d,j}^a-x_{d,j}^b)|+|\mathop{\sum}\limits_{j=1}^Nl_{d,ij}^k(y_{d,j}^a-y_{d,j}^b)|\}\\
    &\leq\mathop{\rm max}\limits_{1\leq i\leq N}\{\mathop{\sum}\limits_{j=1}^N|l_{d,ij}^k(x_{d,j}^a-x_{d,j}^b)|+\mathop{\sum}\limits_{j=1}^N|l_{d,ij}^k(y_{d,j}^a-y_{d,j}^b)|\}\\
    &=\mathop{\rm max}\limits_{1\leq i\leq N}\{\mathop{\sum}\limits_{j=1}^N|l_{d,ij}^k|\cdot(|x_{d,j}^a-x_{d,j}^b|+|y_{d,j}^a-y_{d,j}^b|)\}\\
    &\leq \mathop{\rm max}\limits_{1\leq i\leq N}\mathop{\sum}\limits_{j=1}^N|l_{d,ij}^k|\cdot(\mathop{\rm max}\limits_{1\leq j\leq N}\{|x_{d,j}^a-x_{d,j}^b|+|y_{d,j}^a-y_{d,j}^b|\})\\
    &=\|\bm{I}_N-\epsilon\bm{L}_d^k\|_\infty\|\bm{X}_d^a-\bm{X}_d^b\|_\infty
    \label{pf}
\end{aligned}
\end{equation}
we can see that inequality (\ref{pf}) is scaled for twice. The first scaling is based on the triangle inequality, hence the condition to take the equal sign is $l_{d,ij}^k(x_{d,j}^a-x_{d,j}^b)\geq0$ and $l_{d,ij}^k(y_{d,j}^a-y_{d,j}^b)\geq0$ for every $1\leq j\leq N$. Since $l_{d,ij}^k$ is always non-negative when $0<\epsilon\leq\frac{1}{\|\bm{A}_d^k\|_\infty}$, i.e.,
\begin{equation}
\begin{aligned}
l_{d,ij}^k=
\left\{ 
    \begin{array}{lc}
    1-\epsilon d_{d,i}^k\geq 0,&i=j\\
    \epsilon a_{d,ij}^k\geq 0,&i\neq j
    \end{array}
\right.
\end{aligned}
\end{equation}
we can get the first scaling to be an equality requires
\begin{align}
\left\{ 
    \begin{array}{lc}
    x_{d,j}^a-x_{d,j}^b\geq0\\
    y_{d,j}^a-y_{d,j}^b\geq0
    \end{array}
\right.1\leq j\leq N
\label{c1}
\end{align}

The condition for the second scaling to be equal is 
\begin{align}
    |x_{d,j}^a-x_{d,j}^b|+|y_{d,j}^a-y_{d,j}^b|=\|\bm{X}_d^a-\bm{X}_d^b\|_\infty=C, \forall j
\end{align}
where $C\in\mathbb{R}$ is a constant.

Assuming that the two position matrices $\bm{X}_d^a,\bm{X}_d^b\in\{\bm{X}_d|\sum_{i=1}^Nx_{d,i}=X_{sum},\sum_{i=1}^Ny_{d,i}=Y_{sum}\}$, we can derive that
\begin{equation}
\begin{aligned}
    C&=\frac{1}{N}\mathop{\sum}\limits_{j=1}^N|x_{d,j}^a-x_{d,j}^b|+|y_{d,j}^a-y_{d,j}^b| \\
    &=\frac{1}{N}\mathop{\sum}\limits_{j=1}^N(x_{d,j}^a-x_{d,j}^b+y_{d,j}^a-y_{d,j}^b) \\
    &=\frac{1}{N}(\mathop{\sum}\limits_{j=1}^Nx_{d,j}^a-\mathop{\sum}\limits_{j=1}^Nx_{d,j}^b+\mathop{\sum}\limits_{j=1}^Ny_{d,j}^a-\mathop{\sum}\limits_{j=1}^Ny_{d,j}^b) \\
    &=\frac{1}{N}(X_{sum}-X_{sum}+Y_{sum}-Y_{sum})\\
    &=0
\end{aligned}
\end{equation}
This indicates that $x_{d,j}^a-x_{d,j}^b=0$ and $y_{d,j}^a-y_{d,j}^b=0$ for every $1\leq j\leq N$. Hence, (\ref{dis}) is an equality if and only if $\bm{X}_d^a=\bm{X}_d^b$, and we have
\begin{align}
    d(g_\theta\ast \bm{X}_d^a,g_\theta\ast \bm{X}_d^b)=d(\bm{X}_d^a,\bm{X}_d^b)=0.
\end{align}

At this point, we have proved that for $\forall\bm{X}_d^a,\bm{X}_d^b\in\{\bm{X}_d|\sum_{i=1}^Nx_{d,i}=X_{sum},\sum_{i=1}^Ny_{d,i}=Y_{sum}\}$,
\begin{align}
    d(g_\theta\ast \bm{X}_d^a,g_\theta\ast \bm{X}_d^b)\leq\alpha d(\bm{X}_d^a,\bm{X}_d^b)
\end{align}
always holds for $\alpha\in(0,1)$. This indicates that the bipartite GCO is a contraction operation when $0<\epsilon\leq\frac{1}{\|\bm{A}_d^k\|_\infty}$, and there exists and only exists one position matrix $\bm{\bar{X}}_d=[\bm{\bar{p}}_{d,1},\bm{\bar{p}}_{r_{d,2}},...,\bm{\bar{p}}_{d,N}]^\top$ such that
\begin{align}
    \bm{\bar{X}}_d=(\bm{I}_N-\epsilon\bm{L}_d^k)\bm{\bar{X}}_d=\mathop{\rm lim}\limits_{l\rightarrow\infty}(\bm{I}_N-\epsilon\bm{L}_d^k)^l\bm{X}_d
\end{align}
which is also called the \textit{Banach point}. 

By eliminating the $\bm{\bar{X}}_d$ on both side, we can derive
\begin{align}
    \bm{L}_d^k\bm{\bar{X}}_d=\bm{L}_d^k[\bm{\bar{x}}_d,\bm{\bar{y}}_d]=\bm{0}
\end{align}
where $\bm{\bar{x}}_d$ and $\bm{\bar{y}}_d$ are the column vectors in $\bm{\bar{X}}_d$. This indicates that $\bm{\bar{x}}_d$ and $\bm{\bar{y}}_d$ are eigenvectors of $\bm{L}_d^k$ corresponding to the zero eigenvalue. Note that the number of zero eigenvalues of $\bm{L}_d^k$ equals the number of sub-nets in $G_d^k$, and $S$ zero eigenvalues $\lambda_1=0,\lambda_2=0,...,\lambda_S=0$ correspond to $S$ orthogonal eigenvectors $\{\bm{u}_1,\bm{u}_2,...,\bm{u}_S\}$. Thereby, we can represent $\bm{\bar{x}}_d$ and $\bm{\bar{y}}_d$ as the linear combinations of $\bm{u}_i$, i.e.,
\begin{equation}
\begin{aligned}
    \left\{ 
    \begin{array}{lc}
    \bm{\bar{x}}_d=\mathop{\sum}\limits_{i=1}^S\alpha_i\bm{u}_i\\
    \bm{\bar{y}}_d=\mathop{\sum}\limits_{i=1}^S\beta_i\bm{u}_i
    \end{array}
\right.
\label{cm}
\end{aligned}
\end{equation}
where $\alpha_i,\beta_i\in\mathbb{R}$ are coefficients of the combinations.

When the united MDSG $G_d^k$ is connected, the number of zero eigenvalues of $\bm{L}_d^k$ equals 1. Note that the vector $\bm{1}_N$ must be a eigenvector of $\bm{L}_d^k$ corresponding to the zero eigenvalue, since
\begin{equation}
\begin{aligned}
    \bm{L}_d^k\bm{1}_N=\begin{pmatrix}
    d_{d,1}-\sum_{j=1}^Na_{d,1j}\\
    d_{d,2}-\sum_{j=1}^Na_{d,2j}\\
    \vdots\\
    d_{d,N}-\sum_{j=1}^Na_{d,Nj}
    \end{pmatrix}=\bm{0}=0\cdot\bm{1}_N.
\end{aligned}
\end{equation}

Based on (\ref{cm}), we can get $\bm{\bar{x}}_d=\alpha_1\bm{1}_N$ and $\bm{\bar{y}}_d=\beta_1\bm{1}_N$ with $\alpha_1,\beta_1\neq0$. Since $\bm{\bar{X}}_d\in\{\bm{X}_d|\sum_{i=1}^Nx_{d,i}=X_{sum},\sum_{i=1}^Ny_{d,i}=Y_{sum}\}$, we have
\begin{align}
    \mathop{\sum}\limits_{i=1}^Nx_{d,i}=\mathop{\sum}\limits_{i=1}^N\alpha_1
    \Rightarrow\alpha_1=\frac{1}{N}X_{sum}\\
    \mathop{\sum}\limits_{i=1}^Ny_{d,i}=\mathop{\sum}\limits_{i=1}^N\beta_1
    \Rightarrow\beta_1=\frac{1}{N}Y_{sum}.
\end{align}
Therefore, when the united MDSG $G_d^k$ is connected, $\bm{\bar{X}}_d\equiv[\bm{p}_c,\bm{p}_c,...,\bm{p}_c]^\top$ holds. This completes the proof.

}

 


\bibliographystyle{IEEEtran}
\balance
\bibliography{IEEEabrv,main-text}


 




\vfill

\end{document}